\documentclass[fleqn,usenatbib]{mnras}

\usepackage{newtxtext,newtxmath}
\usepackage[T1]{fontenc}
\usepackage{ae,aecompl}

\usepackage{graphicx}	% Including figure files
\usepackage{amsmath}	% Advanced maths commands
\usepackage{amsfonts}   % Better fonts for equations
\usepackage{amssymb}	% Extra maths symbols
\usepackage{upgreek}    % for \upmu
\usepackage[]{footnote}
\DeclareMathOperator*{\median}{median}

\title[A GPU-accelerated search for pulsars in the HTRU survey]{The High Time Resolution Universe survey XIV: Discovery of 23 pulsars through GPU-accelerated reprocessing}

\author[V.~Morello et al.]{
V.~Morello,$^{1,3}$\thanks{E-mail: vincent.morello@postgrad.manchester.ac.uk}
E.~D.~Barr,$^{2,3}$
S.~Cooper,$^{1}$
M.~Bailes,$^{3,4}$
S.~Bates,$^{1}$
N.~D.~R.~Bhat,$^{5}$ 
\newauthor
M.~Burgay,$^{6}$
S.~Burke-Spolaor,$^{7,14}$
A.~D.~Cameron,$^{2}$
D.~J.~Champion,$^{2}$ 
R.~P.~Eatough,$^{2}$
\newauthor
C.~M.~L.~Flynn,$^{3,4}$
A.~Jameson,$^{3,4}$
S.~Johnston,$^{8}$
M.~J.~Keith,$^{1}$
E.~F.~Keane,$^{9}$
\newauthor
M.~Kramer,$^{1,2}$
L.~Levin,$^{1}$
C.~Ng,$^{10}$
E.~Petroff,$^{11}$
A.~Possenti,$^{6,15}$
B.~W.~Stappers,$^{1}$
\newauthor
W.~van Straten,$^{12}$
C.~Tiburzi$^{13}$
\\
$^{1}$Jodrell Bank Centre for Astrophysics, School of Physics and Astronomy, The University of Manchester, M13 9PL, UK\\ 
$^{2}$Max-Planck-Institut f\"{u}r Radioastronomie, Auf dem H\"{u}gel 69, D-53121 Bonn, Germany. \\
$^{3}$Centre for Astrophysics and Supercomputing, Swinburne University of Technology, Mail H39, PO Box 218, VIC 3122, Australia. \\
$^{4}$ARC Center of Excellence for Gravitational Wave Discovery (OzGrav), Swinburne University of Technology, Mail H11, PO Box 218,\\ VIC 3122, Australia.\\
$^{5}$International Centre for Radio Astronomy Research, Curtin University, Bentley, WA 6102, Australia.\\
$^{6}$INAF - Osservatorio Astronomico di Cagliari, Via della Scienza 5, I-09047 Selargius (CA), Italy.\\
$^{7}$Center for Gravitational Waves and Cosmology, West Virginia University, Chestnut Ridge Research Building, Morgantown, WV 26505,\\ USA.\\
$^{8}$CSIRO Astronomy \& Space Science, Australia Telescope National Facility, P.O. Box 76, Epping, NSW 1710, Australia.\\
$^{9}$SKA Organisation, Jodrell Bank Observatory, SK11 9DL, UK. \\
$^{10}$Department of Physics and Astronomy, University of British Columbia, 6224 Agricultural Road, Vancouver, BC V6T 1Z1, Canada.\\
$^{11}$Anton Pannekoek Institute for Astronomy, University of Amsterdam, P.O. Box 94249, 1090 GE Amsterdam, The Netherlands.\\
$^{12}$Institute for Radio Astronomy \& Space Research, Auckland University of Technology, Private Bag 92006, Auckland 1142, New Zealand.\\
$^{13}$Fakult\"{a}t fur Physik, Universit\"{a}t Bielefeld, Postfach 100131, D-33501 Bielefeld, Germany. \\
$^{14}$Department of Physics and Astronomy, West Virginia University, P.O. Box 6315, Morgantown, WV 26506, USA. \\
$^{15}$Universit\`{a} di Cagliari, Dip. di Fisica, S.P. Monserrato-Sestu Km 0,700 - 09042 Monserrato (CA) - Italy
}

\date{Accepted XXX. Received YYY; in original form ZZZ}

\pubyear{2018} 

\begin{document}
\label{firstpage}
\pagerange{\pageref{firstpage}--\pageref{lastpage}}
\maketitle

\begin{abstract}
We have performed a new search for radio pulsars in archival data of the intermediate and high Galactic latitude parts of the Southern High Time Resolution Universe pulsar survey. This is the first time the entire dataset has been searched for binary pulsars, an achievement enabled by GPU-accelerated dedispersion and periodicity search codes nearly 50 times faster than the previously used pipeline. Candidate selection was handled entirely by a Machine Learning algorithm, allowing for the assessment of 17.6 million candidates in a few person-days. We have also introduced an outlier detection algorithm for efficient radio-frequency interference (RFI) mitigation on folded data, a new approach that enabled the discovery of pulsars previously masked by RFI. We discuss implications for future searches, particularly the importance of expanding work on RFI mitigation to improve survey completeness. In total we discovered 23 previously unknown sources, including 6 millisecond pulsars and at least 4 pulsars in binary systems. We also found an elusive but credible redback candidate that we have yet to confirm.

\end{abstract}


%%%%%%%%%%%%%%%%%%%%%%%%%%%%%%%%%%%%%%%%%%%%%%%%%%
%%%%%%%%%%%%%%%%% BODY OF PAPER %%%%%%%%%%%%%%%%%%
%%%%%%%%%%%%%%%%%%%%%%%%%%%%%%%%%%%%%%%%%%%%%%%%%%

\section{Introduction}

The Southern High Time Resolution Universe (HTRU) project \citep[see][for a complete description]{HTRUI} is an extensive survey of the Southern Sky for pulsars and fast transients performed with the Parkes multibeam receiver \citep{StaveleySmith1996} between 2008 and 2013. The survey area was partitioned into three sections: a low, intermediate and high galactic latitude component referred to as lowlat, medlat and hilat, with on-sky integration times of 70, 9 and 4.5 minutes respectively. Lowlat was designed to discover relativistic binaries in the Galactic plane, medlat to find new bright millisecond pulsars worthy of being included in pulsar timing array experiments, and hilat to improve the pulsar census at high galactic latitudes and explore an under-searched portion of the sky for radio transient events such as fast radio bursts \citep[FRBs;][]{LorimerBurst, Thornton2013}.

Prior to this work, the medlat portion of the survey had already been entirely searched for periodic pulsar signals once \citep{Bates2012, LevinThesis}, while 32\% of hilat had been processed \citep{ThorntonThesis}. Due to limited computing resources, none of these searches were sensitive to pulsars with high orbital acceleration; yet with more than 80\% of the known millisecond pulsar population found in binary systems, running an extensive acceleration search \citep{jk91, Ransom2002} would ensure the fulfillment of the initial science goals of the survey.

This, however, comes with two major obstacles: high data rates and the ubiquitous presence of radio-frequency interference (RFI). The HTRU survey is a prime target to develop and test new software with the goal of demonstrating quasi real-time processing with minimal human intervention in candidate classification, as will be required on upcoming telescopes such as CHIME \citep{PulsarScienceCHIME} or the SKA \citep{PulsarScienceSKA}. Running new analyses of archival data with refined tools can also lead to a significant number of new discoveries even in a previously searched portion of the parameter space, as demonstrated on the Parkes Multibeam Pulsar Survey \citep[PMPS;][]{Manchester2001} by several successful re-processings \citep[e.g.][]{Keith2009, KeanePMPS2010, EatoughPMPS2013, Knispel2013}.

Here we present a new search of the intermediate and high latitude parts of the HTRU survey. In \S 2, we delve into the details of our processing pipeline. We also developed a new efficient RFI mitigation algorithm based on outlier detection, of which we release a fast python implementation with this paper. In \S 3, we briefly present the resulting 23 new pulsar discoveries; full timing solutions will be provided in upcoming publications. We also report a credible accelerating MSP candidate that we failed to confirm despite extensive radio re-observations, and invite the wider community to pursue the effort. In \S 4, we pursue a comparison of our new search pipeline with the one previously used on the same data, and finally discuss in \S 5 the potential implications for future pulsar searches.

%%%%%%%%%%%%%%%%%%%%%%% METHODS SECTION %%%%%%%%%%%%%%%%%%%%%%%%
\section{Methods and data analysis}
\label{sec:methods}

\begin{figure*}
\includegraphics[width=1.00\textwidth]{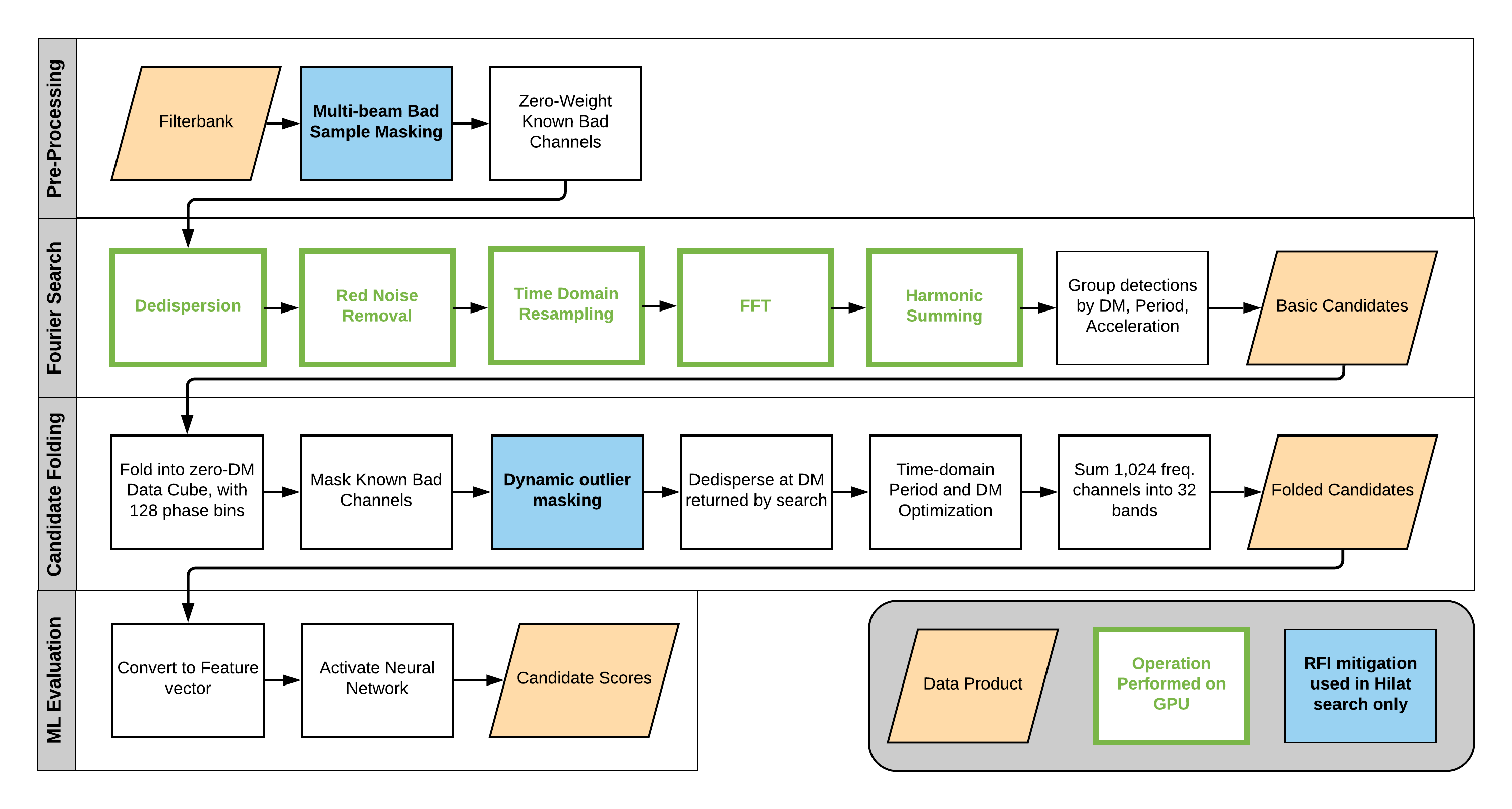}
\caption{Detailed overview of the full processing pipeline. Results from the processing of medlat encouraged the use of extra RFI mitigation layers on the subsequent hilat search, including a new outlier masking algorithm described in \S \ref{sec:candidate_folding}.}
\label{fig:pipeline_diagram}
\end{figure*}

Below we describe the pulsar parameter space searched and the three major operations of the search pipeline: Fourier search, candidate folding, candidate classification along with the RFI mitigation methods employed. The detailed sequence of processing steps is summarized in Fig. \ref{fig:pipeline_diagram}.

\subsection{Search Parameters}
\label{subsec:search_parameters}

The first processing passes of the med- and hilat portions of the HTRU survey used \textsc{SigProc}-based processing pipelines \citep{HTRUI}. While these pipelines were clearly effective, discovering 122 new pulsars, resource-constraints limited these analyses to pulsars without significant acceleration. The development of the GPU-accelerated \textsc{Peasoup}\footnote{\url{https://github.com/ewanbarr/peasoup}} search code removed these constraints allowing for both regions to be searched out to moderate accelerations. The \textsc{Peasoup} pipeline does not offer improved sensitivity for slow or unaccelerated pulsars and as such we selected a parameter space to search that focused on the discovery of binary MSPs. Table \ref{tab:search_parameters} shows the parameters for the searches conducted on medlat and hilat. These parameters are explained throughout this section. For medlat we selected a maximum DM to search of 400 pc cm$^{-3}$. This corresponds to $\sim1$ ms of dispersive smearing within the 391-kHz frequency channels of the \textsc{BPSR} backend \citep[described in \S 3.2 of][]{HTRUI}. For hilat we limited ourselves to twice the maximum line-of-sight DM according to the NE2001 Galactic free electron density model \citep{NE2001} with a cap of 1000 pc cm$^{-3}$.

In both cases trial DMs to search were generated using the method explained at length in \S 2.3 of \citet{LevinThesis}, which we briefly outline here. A pulse of intrinsic width $W_{\rm int}$ is broadened to a total width $W_{\rm tot}$ by three numerical effects, namely sampling, smearing due to intra-channel dispersion, and de-dispersion at an incorrect DM. Each have their associated time scales, respectively $\tau_{\rm samp}$, $\tau_{\rm smear}$ and $\tau_{\rm \Delta DM}$. These effects add in quadrature so that we have

\begin{equation}
W_{\rm tot}^2 = W_{\rm int}^2 + \tau_{\rm samp}^2 + \tau_{\rm smear}^2 + \tau_{\rm \Delta DM}^2.
\end{equation}
The first three terms in the sum are predetermined by the observing setup and by the DM of the source in the case of $\tau_{\rm smear}$. Only $\tau_{\rm \Delta DM}$ can be reduced by adopting a narrower DM trial spacing. We can rewrite the above as 

\begin{equation}
W_{\rm tot}^2 = W_{\rm eff}^2 + \tau_{\rm \Delta DM}^2,
\end{equation}
where $W_{\rm eff}$ is the effective pulse width once all unavoidable broadening effects have been taken into account. The idea is to choose the spacing from one DM trial to the next so that $\tau_{\rm \Delta DM}$ never exceeds some small fraction $\epsilon$ of $W_{\rm eff}$, which can be written:

\begin{equation}
\label{eq:dm_spacing_constraint}
W_{\rm tot} \leq (1 + \epsilon) W_{\rm eff}.
\end{equation}
The constraint above is usually expressed in terms of a so-called \textit{DM tolerance parameter}, defined as $\rm DM_{\rm tol} = 1 + \epsilon$. $\rm DM_{\rm tol} = 1.1$ in all our searches. $W_{\rm int}$ also constitutes a free parameter and represents the minimum intrinsic pulse width one expects in the data.

Due to an oversight\footnote{beam coordinate offsets from the centre beam were occasionally applied in degrees instead of radians, before passing the result as input to the NE2001 model}, approximately $3\%$ of the pointings were searched out to DM values less than the NE2001 model predictions for their lines of sight. These pointings are predominantly located close to the Galactic plane where there exists deeper coverage from lowlat \citep{NgLowlat2015} but also the PMPS \citep{Manchester2001} and SUPERB surveys \citep{SUPERBI}. For both medlat and hilat we selected an acceleration range of $|a| < 50$ m s$^{-2}$. Calculating the maximum l.o.s. acceleration for all the binary pulsars in version 1.58 of the ATNF Pulsar Catalogue \citep{PSRCAT} we find that this range contains $\sim97\%$ of the known pulsar population with most of the remaining pulsars belonging to more massive double neutron star systems less likely to be found far from the Galactic plane. As with the DM trials, the acceleration step size was determined based on an acceleration tolerance parameter $a_{\rm tol}$ of 1.1.

\begin{table}
\centering
\caption{Search parameters for both medlat and hilat searches. $W_{\rm int}$ and DM$_{\rm tol}$ control the spacing of consecutive DM trials (details in \S \ref{subsec:search_parameters}). Likewise, $W_{\rm int}$ and $a_{\rm tol}$ control that of the acceleration trials. $f_{5}$ and $f_{25}$ are parameters of the dereddening of the Fourier spectrum (see \S \ref{subsec:search_code}).}
\label{tab:search_parameters}
\begin{tabular}{lll}
\hline
\hline
Parameter                       & medlat   & hilat\\
\hline
DM$_{\rm max}$ (pc cm$^{-3}$)       & 400      & See text. \\
$N_{\rm DM trials}$                   & 1347     & See text.\\
DM$_{\rm tol}$                    & 1.1      & 1.1\\
$W_{\rm int}$ ($\mu$s)            & 64       & 40\\
$|a_{\rm max}|$ (m s$^{-2}$) & 50 & 50 \\
$\delta a$ (m s$^{-2}$)  & 1.46     & 3.66  \\
$a_{\rm tol}$            & 1.1 & 1.1 \\
$N_{\rm acctrials}$         & 71       & 30 \\
$N_{\rm harmonics}$             & 16       & 16  \\
$f_{5}$ (Hz)          & 0.05     & 0.05  \\
$f_{25}$ (Hz)       & 0.5      & 0.5\\
$f_{\rm min}$ (Hz)                   & 0.1      & 0.1   \\
$f_{\rm max}$ (Hz)       & 1100     & 1100  \\
$N_{\rm beams}$ & 95,940 & 358,644\\ 
Fraction processed & 100\% & 81\%\\
\hline
\end{tabular}
\end{table}

\subsection{Search code overview}
\label{subsec:search_code}

\textsc{Peasoup} implements a time-domain resampling acceleration search \citep{jk91} on \textsc{nvidia} GPUs. The pipeline includes dedispersion through the \textsc{Dedisp} library \citep{Barsdell2012}, dereddening (low-frequency noise removal) in the Fourier domain, resampling, FFT, harmonic summing up to the $16^{\rm th}$ harmonic, peak detection and optional time series folding.

To save computing time, \textsc{Peasoup} performs dereddening only once per DM trial, producing a time series with reduced low-frequency noise that is subsequently searched at all trial accelerations. The dereddening method consists of taking the FFT of the dedispersed time series and scaling the complex-valued Fourier coefficients by an appropriate real-valued, frequency-dependent factor \citep[following \S 3.1 of][]{Ransom2002}, before inverse transforming the corrected Fourier amplitudes back to the time domain. The scaling factor is calculated so that all bins of the corrected power spectrum follow a chi-squared distribution with two degrees of freedom (d.o.f.), i.e.

\begin{equation}
\label{eq:dereddening}
B_i = A_i \sqrt{ \frac{2 \ln 2} {\underset{i-m \leq k \leq i+m}{\mathrm{median}} \{|A_k|^2\}} },
\end{equation}
where the $A_i$ are the complex Fourier amplitudes of the original dedispersed time series, the $B_i$ represent the dereddened amplitudes, the denominator term is a robust estimate of the Fourier power around bin number $i$ and $m$ is half the size of a running median window. The factor $2 \ln 2$ is the expected median value of a chi-squared distribution with two d.o.f. An underlying assumption here is that the expected value of the power spectrum is constant across the median window; if such is the case, then the $B_i$ can be expected to have normally distributed real and imaginary components each with zero mean and unit variance, like the Fourier transform of Gaussian white noise.

While Eq. \ref{eq:dereddening} corresponds to scaling by a running median of width $w = 2 m + 1$, in practice \textsc{Peasoup} breaks the spectrum into blocks of $w=5$ bins, takes the median of each block and then interpolates the output to all frequencies to obtain an \textit{approximate} running median, which is much better suited to parallel architectures. The process can be iterated multiple times (median of medians) to obtain a local Fourier power estimate over windows of width $5^n$. Since large median windows at low frequency tend to underestimate the red noise power, we define two parameters $f_{5}$ and $f_{25}$, that correspond to the frequencies at which we switch from a median smoothing window of 5 to 25 spectral bins and 25 to 125, respectively. Note that since we deredden only once per DM trial, the dereddening applied at zero acceleration must also be valid for higher accelerations. Thus the selection of the $f_{5}$ and $f_{25}$ parameters is important as the size of the smoothing window should be at least twice the expected number of Fourier bins drifted by an accelerated pulsar over the course of an observation. As our smoothing window size is capped at 125 bins we start to see reductions in sensitivity for signals above 1294 Hz at the maximum acceleration of 50 m s$^{-2}$. The reduction in S/N due to this effect is a function of the true S/N of the given signal. We find empirically that signals with true S/N $\lesssim 12$ are unaffected, and stronger ones are never reduced to below S/N $\approx 12$.

Two basic levels of RFI rejection were used in both searches. Firstly 158 frequency channels (15.4 \% of the band) known to be contaminated by narrow-band RFI were ignored during dedispersion. Secondly, a list of Fourier spectral bins known to contain high-occupancy periodic RFI were zero-weighted such that they would not affect the acceleration search. These bins amounted to no more than 0.1\% of the Fourier spectrum. As with dereddening, this process was performed once per DM trial. Finally, for the hilat search one further level of RFI rejection was included. Here the multi-beam RFI detection system of \citet{kbb+12} was used to identify time samples affected by near-field impulsive RFI, which were replaced by representative noise samples drawn from the distribution of surrounding valid data.

\subsection{Processing}
\label{subsec:processing}

The data were processed on the Green II (G2) supercomputer at Swinburne University of Technology\footnote{\url{https://supercomputing.swin.edu.au/}}. G2 is composed of a 3.4-petabyte lustre file system and two computing clusters, gSTAR and SwinSTAR. They host respectively 102 NVIDIA Tesla C2070s (two per node) and 64 NVIDIA Tesla K10 accelerators (one per node, with two GPUs). Processing of the medlat portion of the survey used both gSTAR and SwinSTAR, while we restricted processing of the hilat portion of the survey to SwinSTAR due to reduced availability of the gSTAR cluster. \textsc{Peasoup} is multi-GPU capable and therefore nodes were assigned processing jobs that used both GPUs.

The use of GPU accelerators allowed us to greatly reduce the overall processing time required. The initial processing of medlat, without acceleration trials, consumed $\sim$700,000 CPU-hours spread over the three-year period during which the survey was being observed. Had all the data been available when the initial processing started, it would have taken a minimum of 8 months on the hardware available at the time \citep{LevinThesis}. In contrast, the accelerated reprocessing of medlat presented here took only 3 months ($\sim$15,000 GPU-hours).

Performance logs taken from the hilat reprocessing show median execution times per DM trial of 43 ms for dedispersion and 128 ms for the acceleration search. Including overheads, we find a median execution time of 182 ms per DM trial for the full search (excluding candidate folding, see \S \ref{sec:candidate_folding}). The performance demonstrated here was a driver for the development of the real-time acceleration search deployed as part of the SUPERB survey \citep{SUPERBI}. Detailed performance comparisons between the old and new pipelines are presented in \S \ref{subsec:speed_comparison}.

As can be seen in Table \ref{tab:search_parameters} the total number of beams processed ($N_{\rm beams}$) does not match the numbers presented in Table 1 of \citet{HTRUI}. In the case of medlat we exceed the total number of beams in the survey due to the inclusion of re-observations of RFI-affected pointings and confirmation observations for newly discovered pulsars. In the case of hilat we were limited to the data available on the G2 file system at the time of processing. This constituted 81\% of the full survey.

\subsection{Improved candidate folding}
\label{sec:candidate_folding}
For each candidate identified by the FFT search, the original data need to be phase-coherently folded at the candidate period, in order to obtain more accurate parameters and additional diagnostic information. It is standard practice to partition the input data in equal-sized time segments dubbed \textit{sub-integrations}, and then for each sub-integration, fold every channel individually. The folded output is therefore a three-dimensional array with frequency, time and phase axes. We refer to it as a \textit{data cube} below. The optimal candidate parameters can then be found via a fine grid search over a range of trial dispersion measures and periods centered around the best candidate solution reported by the FFT search.

\subsubsection{The need for additional RFI mitigation}

For our search of medlat we used the \textsc{dspsr} \citep{DSPSR} and \textsc{psrchive} software packages \citep{Hotan2004} to perform respectively candidate folding and time-domain candidate parameter optimization. Our processing strategy was to put the entire burden of RFI rejection on the final candidate classification algorithm (described in \S \ref{subsec:classification}). However we subsequently found that this approach was sub-optimal; for observations affected by strong interference, the optimization process may not converge to the true pulsar parameters. It is only guaranteed to report the parameters that maximize a statistical test for the presence of a pulse in the corresponding integrated profile. Occasionally, RFI distorts this test to an extent such that the final candidate plot is impossible to identify as a pulsar even by a trained eye. The accuracy of the candidate classification stage, regardless of it being human or algorithmic, is dependent on the quality of the candidate plots produced. 

\begin{figure*}
\includegraphics[width=0.88\textwidth]{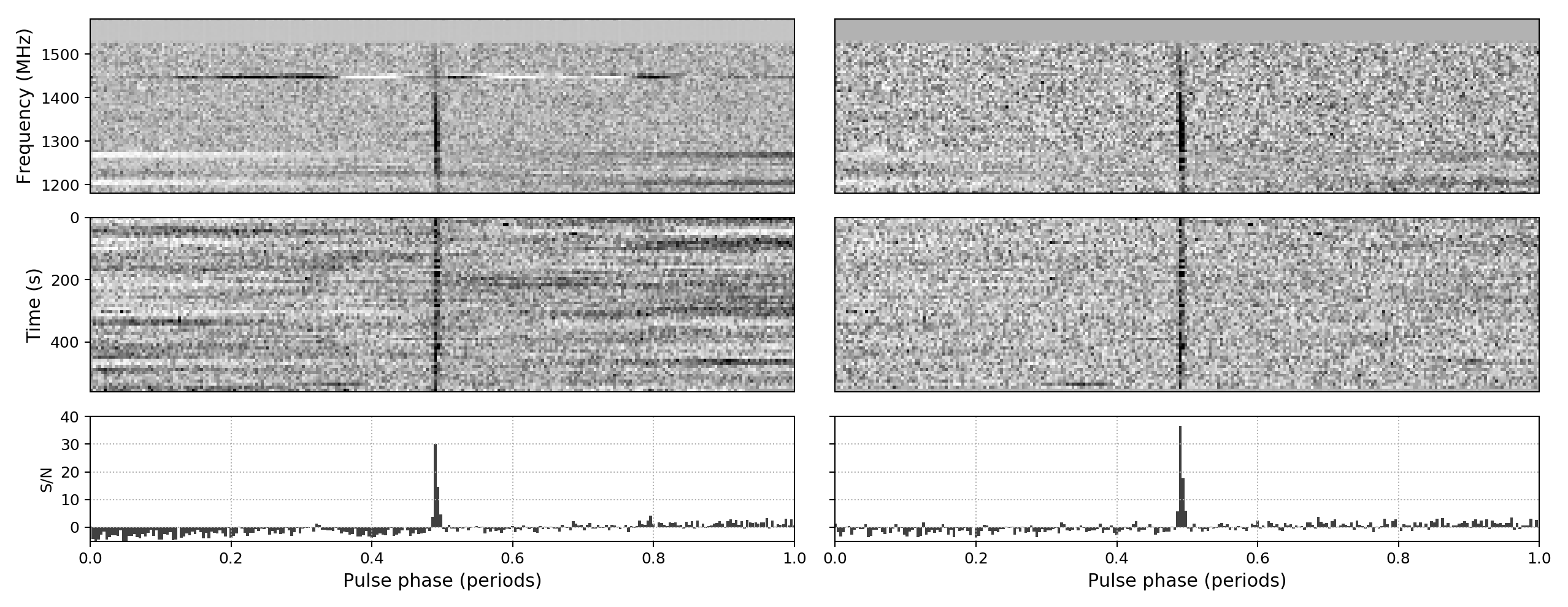}
\caption{A folded 9-minute observation of PSR J0736$-$6304 ($P=4.86$ s, $\mathrm{DM}=19.4~\mathrm{pc~cm}^{-3}$) before and after the RFI mitigation algorithms described in \S \ref{sec:candidate_folding} have been applied (left and right columns respectively). Top: frequency-phase plots, where we have grouped and summed the original 1024 frequency channels into 64 sub-bands for readability. Center: time-phase plots. Bottom: integrated pulse profiles over the entire observation. In both cases, integrated profiles have been normalized by a factor $s \sqrt{N}$, where $N$ is the number of unmasked profiles in the cube, and $s$ the standard deviation of unmasked data. The top 150 frequency channels are permanently occupied by telecommunication signals and masked by an RF filter \citep{HTRUI}. The peak S/N of the pulse increases noticeably from 31 to 37. Improving the general quality of candidate plots reduces the probability of genuine pulsars being improperly rejected in the classification stage of a blind search.}
\label{fig:before_after_cleaning}
\end{figure*}

For the hilat search we therefore decided to develop our own software package to perform candidate folding and optimization called \textsc{cubr} \citep[see Chapter 4 of][]{MorelloMScT}\footnote{\url{hdl.handle.net/1959.3/434704}}, which was also integrated into the SUPERB pulsar search pipeline \citep{SUPERBI}. Its main feature is that it removes interference in the folded data cube before the optimization process. For this, it uses a simple outlier rejection algorithm to identify and suppress data that significantly deviates from the dominant distribution. In what follows, $A$ is the data cube and $i$, $j$ and $k$ indices along the frequency, time and phase axes respectively. $A_{ijk}$ represents a single data point while $A_{ij}$ denotes the pulse profile for channel index $i$ and sub-integration $j$. Before we apply any RFI mitigation, the dispersion measure of the folded data is set to zero, and we subtract from every profile $A_{ij}$ its own median value. It should be noted that RFI comes in many forms; however, by contrast with pulsar emission, it rarely exhibits any dispersion and often occurs in either a narrow band of frequencies or in a short interval of time. These properties guided the design of the method below, which we demonstrate on a folded observation of a known pulsar affected by interference (Fig. \ref{fig:before_after_cleaning}, left panel) to serve as evidence of its effectiveness.

\subsubsection{Step 1: Removal of bad frequency and time intervals}

\begin{figure*}
\includegraphics[width=0.92\textwidth]{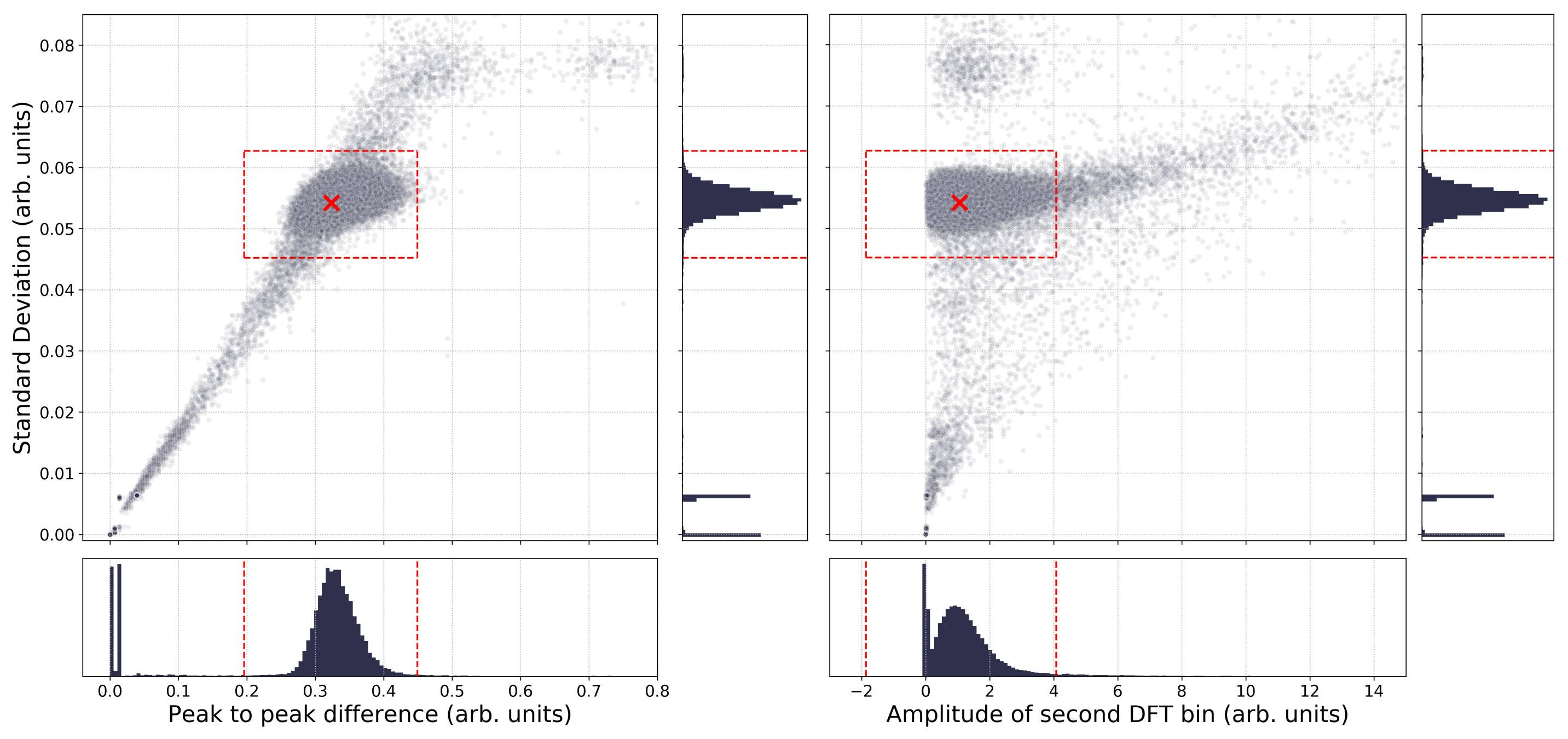}
\caption{Visualization of the outlier rejection algorithm based on Tukey's rule, here applied to the pulsar observation shown in the left panel of Fig. \ref{fig:before_after_cleaning}. For each profile in the data cube we compute three features (see \ref{sec:candidate_folding}), and project all profiles in the resulting three dimensional space of which we show two 2D slices in the scatter plots above. For each dimension of the feature space, we determine an acceptable value range based on the median and inter-quartile range of the whole distribution; their histograms are displayed along scatter plot axes. The resulting decision boundary in feature space is a 3D box whose 2D projections are visible as dashed lines in the scatter plots above.
%The peaks in the histograms correspond to the 150 frequency channels permanently masked at hardware level; profiles from those channels densely cluster around coordinates (0,0,0) here.
}
\label{fig:outlier_rejection}
\end{figure*}

\begin{figure*}
\includegraphics[width=1.00\textwidth]{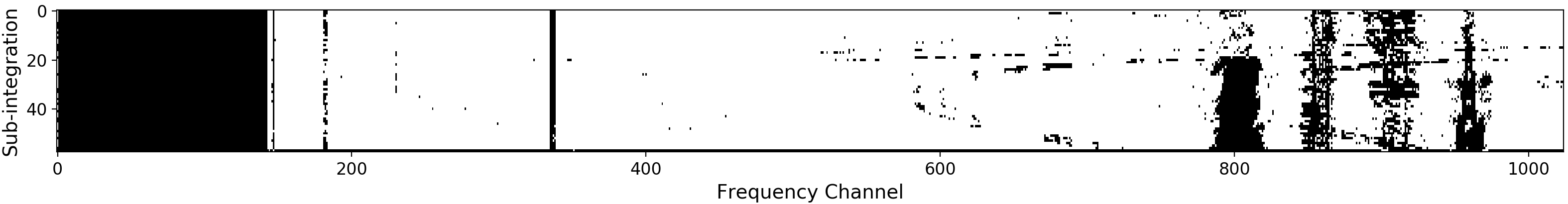}
\caption{Profile mask determined from the analysis shown in Fig. \ref{fig:outlier_rejection}. Each pixel corresponds to a pulse profile, which is part of a given frequency channel (X axis) and sub-integration (Y axis). Outliers appear in black and the corresponding data are zero-weighted before further processing. Bad frequency channels are successfully flagged in this example. Arc-like patterns are visible, revealing narrow-band interference sources whose frequencies change over a timescale of $\approx 10$ seconds.}
\label{fig:profile_mask}
\end{figure*}

Here we flag abnormal profiles $A_{ij}$ based on three numerical profile features:
\begin{enumerate}
\item Standard deviation.
\item Difference between maximum and minimum value (peak-to-peak difference).
\item Absolute value of the second bin of the profile's Fourier transform. This helps identify strong low-frequency noise and signals with a sine-wave shape.
\end{enumerate}
Every profile is associated to a point in this three dimensional feature space (Fig. \ref{fig:outlier_rejection}), which reveals a number of outliers. To identify them, we apply Tukey's rule \citep[]{Tukey1977, Chandola2009} to each feature: a value is considered anomalous if it falls out of the interval $[Q_1 - q R, Q3 + q R]$ where $Q_1$ is the 25th percentile of the distribution, $Q_3$ the 75th percentile and $R = Q_3 - Q_1$ is called the interquartile range. $q$ is a free parameter that can be accurately mapped to a false rejection probability if most of the data follow a normal distribution. Tukey's original recommendation is $q=1.5$; we set $q=2.0$ in our processing following tests on a sample of RFI contaminated pulsar observations. $Q_1$, $Q_3$ and $R$ are not distorted by the presence of outliers, which make them a good choice as opposed to the mean or the standard deviation. A profile that stands out as anomalous with respect to any of the three features above is flagged as bad (Fig. \ref{fig:profile_mask}) and zero-weighted before further processing.

\subsubsection{Step 2: Removal of zero-DM interference pulses}

Next, we identify short duration bright interference bursts that frequently appear in Parkes data. These are not dispersed, so they are more easily identified in the so-called sub-integrations plot, i.e. the sum of the data cube along the frequency axis that we denote $I_{jk}$:

\begin{equation}
I_{jk} = \sum_{i=1}^{n_c} A_{ijk},
\end{equation}
where $n_c$ denotes the number of frequency channels, 1024 in the case of HTRU data. A short interference burst manifests itself in a single sub-integration as a small interval of phase bins with an abnormally high value. We may therefore identify them by separately applying Tukey's rule to every column of $I$, where a given column corresponds to a constant phase. For a bad data point $(j_0, k_0)$ thus flagged in $I$, we replace the associated data $A_{i j_0 k_0}$ across the frequency dimension of the cube according to

\begin{equation}
A_{i j_0 k_0} = \frac{1}{n_c} \median_{1 \leq j \leq n_t} \{I_{j k_0}\},
\end{equation}
where $n_t$ is the number of sub-integrations, which was set to 32 in our processing. The replacement is performed for every $i$ from $1$ to $n_c$. The overall effect of the application of the cleaning algorithms described here can be seen in the right panel of Fig. \ref{fig:before_after_cleaning}. Once RFI mitigation has been applied, we proceed with a standard optimization stage and produce a candidate plot to be evaluated in the final classification stage. 

\subsubsection{Possible limitations}

The upside of our proposed method lies in that it requires no explicit manual parametrization of RFI, having only a single free parameter $q$ defined above. Instead, interference is defined as a small subset of the data cube that significantly differs from the rest, a flexible approach that should in principle work in any RFI environment. We do however make a few implicit assumptions about what the properties of a pulsar signal should be: broadband, dispersed, with no pulses vastly brighter than the others. Should any of these assumptions not hold, there is a risk of erasing a significant part of a pulsar's signal, but that situation should occur infrequently. For the first step to remove a pulse of a genuine pulsar source in a given channel and sub-integration, it should at least be distinguishable from white noise; however, its signal-to-noise ratio (S/N) can be expected to be $\sqrt{n_c \times n_t} \simeq 181$ times lower than the S/N of the pulse integrated over the whole band and observation length. The second step can potentially remove sub-integration samples containing bright individual pulses of an intermittent or nulling pulsar with low DM. Such sources should be more easily detected in a single pulse search however, which are now routinely performed on most pulsar surveys including HTRU \citep{BurkeSpolaor2011}.

\subsubsection{Code release}

A fast, standalone Python implementation of the two RFI mitigation algorithms described above is made publicly available\footnote{\url{https://github.com/v-morello/clfd}}. It comes with an interface to the \textsc{psrfits} format, via the Python bindings of the \textsc{psrchive} package. We also note that it can be readily applied to perform RFI excision on pulsar timing data.

\subsection{Candidate classification}
\label{subsec:classification}

Most large scale pulsar searches of the past two decades have produced candidate numbers large enough that visual inspection of the entire search output is not a reasonable option \citep{Lyon2016}. Our re-processing was no exception and left us with 17.6 million candidates to sort through. Furthermore, medlat had been entirely searched once before, making the number of expected discoveries lower than on fresh data and the cost effectiveness of extensive visual inspection worse. Developing an accurate Machine Learning (ML) classifier was therefore a requirement to make our search successful.

Our classification algorithm SPINN consists of a simple two-layer artificial neural network (ANN). Its first version (SPINN v1 hereafter) has previously been described at length in \citet{Morello2014}, and was used to evaluate candidates from the medlat portion of the survey. The changes subsequently made to the folding software and candidate format prior to running the hilat search (\S \ref{sec:candidate_folding}) were also an opportunity to make adjustments to the classification stage of the pipeline, resulting in an improved SPINN v2. In this section we will limit ourselves to a description of the design guidelines we followed for SPINN v2 and an evaluation of its classification accuracy and speed. Further details and an in-depth explanation of ANNs can be found in Chapter 3 of \citet{MorelloMScT}.

\subsubsection{Candidate features}

In any Machine Learning (ML) problem, each data instance must be converted into a set of descriptive numerical properties called features, that are then fed as an input vector to the ML algorithm. The choice of features plays a critical role in the algorithm's success. For the design of features we followed the general guidelines below:

\begin{itemize}
\item Keep the feature count below 10. An important constraint encountered in pulsar candidate classification is the low number of available real pulsar training examples, on order of 1,000 at best for a given instrument. Intuitively, these can only sparsely sample a 10-dimensional feature space, leaving vast volumes where the classification function to be learned is undefined, but where unseen real pulsar examples may nonetheless lie. This is an aspect of the so-called \textit{curse of dimensionality} \citep[e.g.][]{StatLearning}. Some learning algorithms such as decision trees and forests \citep{Breiman2001} are intrinsically robust to high dimensional data because they operate only on select subsets of the feature space, but such is not the case of ANNs.

\item Choose only features with high relevance, i.e. highly correlated to the class labels (but not necessarily in a linear sense). This is in line with properly spending the limited feature budget we imposed above.

\item Prefer monotonic features. A monotonic feature is such that an increase (or decrease) in its value always corresponds to an increased probability of positive data class membership. When using only monotonic features, both the positive and negative data classes tend to lie in distinct corners of the feature space, which means that the optimal decision boundary between both classes has a simple parametrization. We found monotonic features to facilitate the training process of our ANN and help to increase the accuracy of its predictions.
\end{itemize}

The list of features we use as inputs to SPINN v2 can be found in Table \ref{tab:features}. We relied heavily on domain knowledge to design them, but we note that more systematic approaches can be adopted. Given a large feature set on a learning problem where high dimensionality is undesirable, a nearly optimal subset of useful features can be determined using so-called feature selection algorithms. Section 5 of \citet{Lyon2016} covers in detail the topic of feature selection in the context of pulsar candidate classification.

\subsubsection{Classifier evaluation}

A classifier's performance is measured by deploying it on a test sample of labeled data that was \textit{not} seen during training. The total number of successful or failed predictions on each data class, positive and negative, can be conveniently tracked in a so-called confusion matrix (Table \ref{tab:confusion_matrix}) from which several figures of merit can be derived. In the case of pulsar candidate classification, there are two simultaneous goals: maximize discovery potential while significantly reducing the number of false positive candidates to be viewed by human operators. An inevitable trade-off exists between the two, that can only be fully captured by measuring the two types of error rates:

\begin{equation}
\begin{split}
\mathrm{FNR} &= \frac{\mathrm{FN}}{\mathrm{FN} + \mathrm{TP}} \\
\mathrm{FPR} &= \frac{\mathrm{FP}}{\mathrm{FP} + \mathrm{TN}}
\end{split}
\end{equation}
where FNR is the false negative rate, here the fraction of pulsars missed, and FPR the false positive rate or  fraction of spurious candidates incorrectly reported as pulsars. In place of FNR we may also use \textit{Recall}, defined as the fraction of positives properly identified:
\begin{equation}
\mathrm{Recall} = \frac{\mathrm{TP}}{\mathrm{FN} + \mathrm{TP}} = 1 - \mathrm{FNR}.
\end{equation}

We trained SPINN v2 on a sample of labeled candidates from medlat and tested it on all the candidates produced by searching the first year of observations of the recent SUPERB survey. SUPERB also uses the Parkes multibeam receiver with the same backend, similar integration times on the sky, and the raw data were processed with the same infrastructure described in sections \ref{subsec:search_code} and \ref{sec:candidate_folding}. Most importantly, SUPERB covers a region of the sky that does not overlap with medlat by design, ensuring the fairness of the test. The test set contained 139 known pulsar observations and 1,418,598 negatives, whose class labels were determined by a combination of cross-referencing with the ATNF pulsar catalogue \citep{PSRCAT} and visual inspection for all potential positives. All known pulsars were thus identified; a few undiscovered pulsars may have remained in the set of negatives, but the potential impact on the measured error rates is negligible. Every test candidate was assigned a score between 0 and 1 by SPINN v2, where a higher score indicates a higher likelihood of being a pulsar. This allowed us to evaluate classification error rates as a function of score threshold. The results can be seen in Fig. 8 of \citet{SUPERBI}. The balance between recall and false positive rate is left to the user; for example, one can expect to find 99\% of known pulsars while still dividing the size of the metaphorical haystack of spurious candidates by a factor of nearly 1,000. 

\begin{table}
\caption{A confusion matrix tracks the four possible outcomes of binary classification on a dataset where the true class labels (noted $+$ and $-$) are known.}
\label{tab:confusion_matrix}
\begin{tabular}{ccc}
                      & \textbf{Predicted $+$} &  \textbf{Predicted $-$} \\
\hline
\textbf{Actual $+$}   & True  Positive (TP)    &  False Negative (FN) \\
\textbf{Actual $-$}   & False Positive (FP)    &  True Negative (TN)  \\
\hline
\end{tabular}
\end{table}

As for classification speed, one instance of SPINN v2 runs on a single CPU and scores approximately 180 candidates per second on an Intel Xeon E5 2.2GHz CPU. This takes into account the time required to read the candidate from disk, compute the features, activate the neural network to obtain a score and store the results. Any classification problem can be trivially parallelized by deploying multiple identical classifier instances. The G2 supercomputer allowed us to run 64 SPINN v2 instances at a time, which means that the entire output of the pulsar searches described here (17.6 million candidates) could be processed in less than 30 minutes. We decided for both searches to visually inspect candidates with a score above a threshold corresponding to 99\% recall; this returned less than 0.1\% of the total output of the search (Table \ref{tab:inspection_summary}). The detailed visual inspection of these high-scoring candidates from the hilat portion of the survey required less than 30 person-hours.

\begin{table}
\caption{List of candidate features used as an input to SPINN v2, used on the hilat search. The formulas used to compute these features can be found in Chapter 3 of \citet{MorelloMScT}.}
\label{tab:features}
\begin{tabular}{r|l|p{5.8cm}|}
\hline
\hline
No. & Name             & Short description \\
\hline
1 & \texttt{snr}       & Signal-to-noise ratio of the integrated profile \\
2 & \texttt{deq}       & Equivalent duty cycle of the integrated profile \\
3 & \texttt{cider}     & Confidence index in true source DM being higher than 2.0 $\mathrm{pc~cm}^{-3}$ \\
4 & \texttt{cscore}    & Level of presence of candidate period in other receiver beams during the same observation \\
5 & \texttt{actime}    & Average correlation coefficient between integrated profile and individual sub-integrations \\
6 & \texttt{acfreq}    & Average correlation coefficient between integrated profile and individual sub-bands \\
7 & \texttt{zratio}    & Ratio between profile S/N at best DM and at DM=0 \\ 
\hline
\end{tabular}
\end{table}

\begin{table*}
\caption{Candidate classification and inspection summary. In both searches, candidates were selected for visual inspection down to a score threshold corresponding to a recall of 99\%.}
\label{tab:inspection_summary}
\begin{tabular}{r|r|r|r|r|r|r}
\hline
\hline
Search & Candidates & Classifier & Recall   & Candidates &  Fraction   & Discoveries \\
       &            &            &          & Inspected  &  Inspected  &  \\
\hline
medlat &  4,375,642 & SPINN v1   & 99\%       &  4,909     & 0.11\%    & 8 \\
hilat  & 13,240,927 & SPINN v2   & 99\%       & 11,849     & 0.09\%    & 14 \\
\hline
\end{tabular}
\end{table*}

%%%%%%%%%%%%%%%%%%%%%%% DISCOVERIES SECTION %%%%%%%%%%%%%%%%%%%%%%%%
\section{Discoveries}
\label{sec:discoveries}

We have found a total of 23 new, confirmed pulsars whose discovery parameters are presented in Table \ref{tab:discovery_params}. Here we only provide positions of the discovery beam and spin periods to six significant figures, leaving full timing solutions to future publications (Keith et al., in prep., Barr et al., in prep.). The observation IDs in which these pulsars were found and the date of their initial identification as candidates worthy of confirmation can be found in Table \ref{tab:discovery_files}. While none of the pulsars reported here have been previously published formally in the literature, we acknowledge that 5 have been independently found by other survey teams and listed on their official websites. Of the 23 confirmed discoveries, 6 are millisecond pulsars (MSPs) and another one (PSR J1517$-$32) appears to be mildly recycled. At least 4 are part of binary systems: PSRs J0636$-$3044, J1517$-$32, J1705$-$1903 and J1754$+$0032.

\begin{table*}
\centering
\caption{Confirmed pulsar discoveries. Where no DM-dependent distance estimation is given, the pulsar's DM lies beyond the limits of the Galaxy according to that model, which likely points at model flaws for some lines of sight. Some of these pulsars, although not formally published, have been independently found by other survey teams.}
\label{tab:discovery_params}
\begin{tabular}{lcccclccccc}
\hline
\hline
PSR J                    & R.A.        & Decl.        & $l$          & $b$            & {$P_0$}     & DM             & S/N& $D_{\rm NE2001}$ & $D_{\rm YMW16}$ & Indep.\\
                         & (hh:mm:ss)  & (dd:mm:ss)   & ($^{\circ}$) & ($^{\circ}$)   & {(s)}       & (pc cm$^{-3}$) &    & (kpc)            & (kpc)           & Discovery\\
\hline
J0125$-$23$^{\clubsuit}$ & 01:25:10    & $-$23:27:26  & 189.1        & $-$81.5        & 0.00367586  & 10             & 33 & 0.4              & 0.9             & G \\
J0636$-$3044$^{\dagger}$ & 06:37:09    & $-$30:50:42  & 239.7	     & $-$16.3        & 0.00394576  & 16             & 11 & 1.0              & 0.7             & \\
J0753$-$0816$^{\dagger}$ & 07:53:33    & $-$08:19:35  & 227.6        & 9.8            & 2.09362     & 38             & 15 & 1.9              & 1.7             & \\
J0839$-$66               & 08:39:37	   & $-$66:34:52  & 281.4        & $-$14.9        & 0.449339    & 83             & 15 & 3.2              & 0.4             & \\
J0951$-$71               & 09:51:24    & $-$71:19:01  & 289.5        & $-$13.3        & 0.212378    & 77             & 16 & 2.5              & 1.7             & \\
J1000$+$08               & 10:00:38	   & $+$08:19:58  & 229.7        & 45.5           & 0.440372    & 21             & 16 & 0.9              & 1.7             & L \\
J1403$-$0314$^{\dagger}$ & 14:03:41	   & $-$03:15:28  & 335.6        & 55.0           & 0.362634    & 31             & 21 & --               & --              & \\
J1517$-$32$^{\clubsuit}$ & 15:17:01    & $-$32:26:05  & 335.4        & 21.1           & 0.0644019   & 26             & 36 & 1.0              & 0.9             & G \\
J1558$-$67               & 15:58:21    & $-$67:35:59  & 319.3        & $-$10.9        & 0.267268    & 105            & 10 & 2.7              & 5.1             & \\
J1654$-$26               & 16:54:39    & $-$26:36:18  & 355.5        & 10.6           & 1.62373     & 129            & 18 & 3.6              & 11.5            & \\
J1703$-$18               & 17:03:26    & $-$18:48:54  & 3.1          & 13.6           & 1.27024     & 46             & 12 & 1.4              & 1.4             & \\
J1705$-$1903$^{*}$       & 17:05:27	   & $-$19:08:04  & 3.1	         & 13.0           & 0.00248022  & 58             & 16 & 1.7              & 2.4             & \\
J1708$+$02               & 17:08:39    & $+$01:48:06  & 22.3         & 23.6           & 0.410772    & 29             & 24 & 1.3              & 1.5             & \\
J1754$+$0032$^{*}$       & 17:54:52	   & $+$00:36:33  & 26.9	     & 12.8           & 0.00441080  & 70             & 10 & 2.4              & 3.7             & \\
J1804$-$2858$^{*}$       & 18:04:19	   & $-$28:58:31  & 2.0	         & $-$3.6         & 0.00149268  & 232            & 11 & 4.9              & 8.8             & \\
J1842$-$27               & 18:42:02    & $-$27:58:19  & 6.7          & $-$10.5        & 0.815269    & 62             & 11 & 1.9              & 2.6             & \\
J1843$-$40               & 18:43:03    & $-$40:33:51  & 355.0        & $-$15.8        & 0.324187    & 66             & 16 & 1.8              & 3.9             & \\
J1921$-$05               & 19:21:04    & $-$05:23:08  & 31.5         & $-$9.0         & 2.22759     & 98             & 16 & 3.2              & 5.7             & G \\
J1940$+$0239$^{\dagger}$ & 19:40:33	   & $+$02:36:41  & 40.9         & $-$9.7         & 1.23224     & 90             & 26 & 3.5              & 5.7             & G \\
J1942$+$0147$^{\dagger}$ & 19:42:27	   & $+$01:50:11  & 40.5         & $-$10.5        & 1.40504     & 151            & 15 & 7.0              & --              & \\
J1947$-$18               & 19:47:31    & $-$18:58:12  & 21.5         & $-$20.6        & 0.00260323  & 25             & 16 & 1.0              & 1.1             & \\
J2228$-$65               & 22:28:18    & $-$65:11:47  & 323.5        & $-$45.8        & 2.74598     & 36             & 31 & 1.8              & --              & S \\
J2354$-$22$^{\clubsuit}$ & 23:54:26    & $-$22:51:53  & 48.1         & $-$76.4        & 0.557996    & 8              & 13 & 0.4              & 0.9             & G \\
\hline
\end{tabular}
\\
\begin{flushleft}
{\footnotesize ${\dagger}$ Timing solutions will be presented in Keith et al. (in prep.)}\\
{\footnotesize ${*}$ Timing solutions will be presented in Barr et al. (in prep.)} \\
{\footnotesize $\clubsuit$ Date of identification in other survey predates ours} \\
G: Green Bank Northern Cap Celestial survey, \url{http://astro.phys.wvu.edu/GBNCC/} \\
S: SUPERB survey, \url{https://sites.google.com/site/publicsuperb/discoveries} \\
L: LOTAAS survey, \url{http://www.astron.nl/lotaas/}
\end{flushleft}
\end{table*}

\begin{table*}
\centering
\caption{Pointing and beam number for each of the newly discovered pulsars. Pointings are uniquely identified by their start UTC date and time. Discovery date refers to the first visual identification as a candidate worthy of re-observation.}
\label{tab:discovery_files}
\begin{tabular}{lcccc}
\hline
\hline
PSR          & Survey & Observation UTC     & Beam & Discovery date\\
             &        & (Y-m-d H:M:S)       &      & (Y-m-d) \\
\hline
J0125$-$23   & H      & 2010-12-31 07:13:55 & 6    & 2015-05-07   \\
J0636$-$3044 & H      & 2011-06-26 00:01:54 & 13   & 2015-05-07   \\
J0753$-$0816 & H      & 2013-01-24 10:57:05 & 3    & 2015-06-11   \\
J0839$-$66   & H      & 2012-03-26 09:31:39 & 10   & 2015-05-28   \\
J0951$-$71   & M      & 2008-12-10 20:37:06 & 1    & 2014-01-07   \\
J1000$+$08   & H      & 2013-12-28 17:40:54 & 3    & 2015-05-28   \\
J1403$-$0314 & H      & 2013-12-30 22:21:55 & 3    & 2015-05-28   \\
J1517$-$32   & H      & 2013-04-05 12:48:51 & 8    & 2015-05-07   \\
J1558$-$67   & M      & 2010-01-21 22:22:30 & 5    & 2014-01-07   \\
J1654$-$26   & M      & 2008-11-23 23:33:40 & 1    & 2014-01-07   \\
J1703$-$18   & M      & 2009-08-21 10:26:40 & 8    & 2014-02-23   \\
J1705$-$1903 & M      & 2010-04-21 16:56:27 & 10   & 2013-10-14   \\
J1708$+$02   & H      & 2013-12-05 23:27:05 & 4    & 2015-05-28   \\
J1754$+$0032 & M      & 2009-08-19 08:37:33 & 12   & 2014-01-16   \\
J1804$-$2858 & M      & 2009-03-08 01:45:33 & 13   & 2013-11-02   \\
J1842$-$27   & M      & 2008-11-24 06:06:51 & 8    & 2014-02-19   \\
J1843$-$40   & H      & 2011-12-08 07:23:15 & 3    & 2015-05-28   \\
J1921$-$05   & H      & 2012-03-31-00:30:01 & 12   & 2015-06-11   \\
J1940$+$0239 & H      & 2013-05-01 17:13:28 & 7    & 2015-06-11   \\
J1942$+$0147 & H      & 2013-10-17 10:30:42 & 6    & 2015-06-11   \\
J1947$-$18   & H      & 2013-07-12 18:01:34 & 4    & 2015-05-07   \\
J2228$-$65   & H      & 2012-01-29 05:18:35 & 1    & 2015-06-11   \\
J2354$-$22   & H      & 2009-04-16 21:19:11 & 13   & 2015-05-28   \\
\hline
\end{tabular}
\end{table*}

\textbf{PSR J1705$-$1903} is a millisecond pulsar with a spin period of 2.48 ms and a very low mass companion (0.047 $\rm{M_{\odot}}$ median mass) indicative of a black widow. It undergoes eclipses for about 15\% of its orbit at 1400 MHz and has an extremely narrow pulse width of $\approx 40 \mu$s. Its usefulness as a timing array pulsar might be hindered by the presence of long-term timing noise likely caused by a non-uniform density of ablated companion material dispersed in its orbital plane \citep[e.g.][]{Fruchter1988b}. However, its narrow pulse width gives it excellent short-term timing properties which can be leveraged to accurately measure DM and scattering variations throughout its orbit; this can in turn be used to characterize the eclipses and the ablation process the companion is undergoing, following work previously done on other similar black widow systems such as PSR J2051$-$0827 \citep{Stappers2001}, PSR J1544$+$4937 \citep{Bhattacharyya2013} or PSR J1810$+$1744 \citep{Polzin2018}.

\textbf{PSR J1804$-$2858} has a spin frequency of 670 Hz, the third highest currently known. Despite being isolated, it shows a positive $\dot{P}$ that is likely caused by acceleration in the local Galactic potential; the possibility of a multi-year eccentric orbit around a massive companion cannot be ruled out however, as was recently demonstrated in the case of PSR J2032$+$4127 \citep{Lyne2015}. PSR J1804$-$2858 has the second highest DM/P ratio of any known pulsar, behind that of the globular cluster pulsar PSR J1748-2446ad \citep{Hessels2006}. The highest DM/P a survey finds has previously been put forward as a measure of survey depth \citep[e.g.][]{Lazarus2013}, which in the case of MSPs is limited by so-called DM smearing: the artificial pulse widening caused by the uncorrected dispersion delay $\Delta t$ within a single frequency channel, given by

\begin{equation}
\label{eq:dm_smearing}
\Delta t = 8.3 \times 10^3 \left( \frac{\mathrm{DM}}{\mathrm{pc~cm}^{-3}} \right) \left( \frac{\Delta \nu}{\mathrm{MHz}} \right)  \left( \frac{\nu}{\mathrm{MHz}} \right)^{-3} \mathrm{s},
\end{equation}
where DM is the dispersion measure of the source, $\Delta \nu$ the width of a frequency channel and $\nu$ the central observing frequency. For the HTRU survey observing setup $\Delta t = 285~\mathrm{\mu s}$ at the pulsar's $\mathrm{DM} = 232~\mathrm{pc~cm}^{-3}$, compared to $2.2$ ms for the previous generation of highly successful Parkes surveys \citep[e.g.][]{Manchester2001}. The increased frequency resolution of HTRU made this discovery possible, and pursuing this trend of finer channelization on the next generation of telescopes will push the MSP detectability horizon further into the Galactic plane, unless limited by interstellar scattering.

\begin{figure*}
\includegraphics[width=0.7\textwidth]{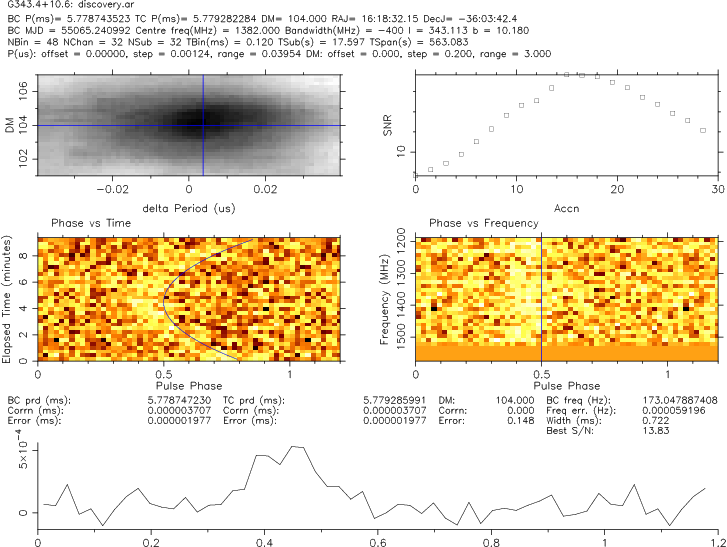}
\caption{Candidate plot for J1618$-$36, generated with \textsc{psrchive}. The phase vs time plot (center left) was generated without correcting for acceleration to make the residual quadratic phase drift visible; the over-plotted thin blue line shows the best-fit quadratic phase model. The upper right plot shows folded signal-to-noise ratio as a function of trial acceleration, giving a best-fit value between 15 and 18 $\mathrm{m~s^{-2}}$ consistent with the 16.3 $\mathrm{m~s^{-2}}$ reported by the Fourier search.}
\label{fig:J1618_pdmp}
\end{figure*}

Finally, one MSP candidate deserves mention. J1618$-$36 was found in the medlat portion of the survey with a period of 5.78 ms and an unusually high acceleration ($16.3~\mathrm{ms}^{-2}$) for a millisecond pulsar. The parameters of its original detection can be found in Table \ref{tab:trollface_params}, and the candidate plot in Fig. \ref{fig:J1618_pdmp}. We note that the positive sign on the acceleration means that the source is accelerating away from the observer, and would correspond to the near side of an orbit. The detection is not due to random noise fluctuations (S/N = 14), and has all the properties that indicate a pulsar nature; it was detected in a single beam, is broadband with a very significantly non-zero dispersion measure and all candidates reported with a similar score by our classifier are genuine pulsars. Its period also lies within an RFI-quiet range. We have unfortunately failed to detect it again after spending a cumulated 6 hours of Parkes time over the years 2015$-$2017, mostly in 9 and 18-minute integrations.

To constrain the possible nature of J1618$-$36, we need to place its acceleration at the time of detection in a broader context. We calculated\footnote{\url{https://github.com/ewanbarr/pyorbit}} the maximum line of sight accelerations (l.o.s.a.) of all binary pulsars for which orbital parameters are available in the ATNF pulsar catalogue. Defining MSPs as pulsars with spin periods below 20 ms, we find that the only known MSP to reach a maximum l.o.s.a. higher than 16.3~$\mathrm{ms}^{-2}$ is PSR J2215+5135 \citep{Hessels2011}, a redback in a 0.17-day orbit around a main-sequence companion star with a median mass of $0.24~\mathrm{M_{\odot}}$. The top six millisecond pulsars ranked by maximum acceleration all have companions with $\geq 0.1~\mathrm{M_{\odot}}$ and orbital periods shorter than a day. Combined with the sequence of non-detections, this tentatively suggests that J1618$-$36 is a redback eclipsed for the vast majority of its orbit, or a new specimen of transitional millisecond pulsar \citep[see e.g.][and references therein]{Campana2018}. If such is the case, it should be detectable as a low mass X-ray binary when in an accreting state. There are two faint ROSAT X-ray sources within one Parkes beam error radius around the discovery position, namely 1RXS J161835.3$-$360329 and 1RXS J161820.8$-$360306 whose angular separation from the candidate position are 41 and 142 arcseconds respectively. Both sources, if placed at a distance of 2.7 kpc, would have 0.5$-$10 keV X-ray fluxes of $\approx 10^{33}$ erg/s, which would be broadly consistent with a transitional MSP hypothesis. We were granted a 2 ks Swift observation (Observation ID: 00033792002, taken on May 28th 2015) to put this idea to the test. If the flux of the source matched that described above then a 46-count source would be detected, but unfortunately no significant source was visible in the field. In spite of the elusiveness of J1618$-$36, we encourage further confirmation attempts and checking of any archival data available due to the very high quality of the initial radio detection, which remains very unlikely to be explained by radio frequency interference or noise.

\begin{table}
\centering
\caption{Original detection parameters of the highly accelerating MSP \textit{candidate} J1618$-$36. The FWHM of a beam of the Parkes multibeam receiver is 14 arcminutes.}
\label{tab:trollface_params}
\begin{tabular}{ll}
\hline
\hline
$\alpha$ (J2000)    & 16h18m32.1s \\
$\delta$ (J2000)    & $-$36d03m42s \\
$l$                 & 343.11$^{\circ}$ \\
$b$                 & 10.18$^{\circ}$ \\
MJD                 & 55065.240992 \\
$P_{\mathrm{bary}}$ & 5.778747 ms \\
DM                  & 104 $\mathrm{pc~cm}^{-3}$ \\
Acc                 & 16.3 $\mathrm{ms}^{-2}$ \\
S/N                 & 14 \\
$D_{\rm NE2001}$    & 2.7 kpc \\
\hline
\end{tabular}
\end{table}

\section{Comparison with previous searches of the same data}

Pulsar searches can be evaluated using three major criteria, namely speed, completeness, and human effort involved in candidate classification. In this section, we compare the previously used \textsc{sigproc}-based pipeline and candidate classification tools with the new pipeline presented in \S \ref{sec:methods}. We will mostly focus on the case of hilat, the survey section on which we used our new RFI mitigation algorithm (see \S \ref{sec:candidate_folding}).

\subsection{Speed}
\label{subsec:speed_comparison}

The speed evaluation was focused on estimating how long it would have taken to perform a complete acceleration search of hilat with the old pipeline, using the same search parameters we selected (Table \ref{tab:search_parameters}) and the same computing cluster. The old pipeline was purely CPU-based. It performed brute-force dedispersion with the \texttt{dedisperse\_all} routine and periodicity search with \texttt{seek}, both of which are part of \textsc{sigproc}. Candidate summary plots were produced with the \textsc{dspsr} \citep{DSPSR} and \textsc{psrchive} \citep{Hotan2004} software packages, using \texttt{dspsr} for folding, \texttt{paz} for masking a static list of bad channels, and finally \texttt{pdmp} for candidate optimization. For the sake of this comparison, the programs that could only use a single CPU core (\texttt{seek} and \texttt{pdmp}) were parallelized using the \texttt{multiprocessing} python library, with negligible overhead.

We measured the execution times of every major pipeline step on G2 (see \S \ref{subsec:processing}), allocating to each of them 3 Intel Xeon E5-2660 CPU cores and 1 Tesla K10 accelerator to process a single 270-second hilat beam. This resource allocation simulated optimal cluster usage conditions in a large scale search, since users are simultaneously limited to a maximum of 64 concurrent jobs, 64 GPUs and $192$ CPU cores. Based on the average number of DM trials in our own hilat processing, we performed the benchmark with 650 DM trials, 30 acceleration trials and 50 candidates folded per beam. The benchmark results are shown in Table \ref{tab:pipeline_runtimes}. We note that under the ``Search" operation are included red noise removal, time domain re-sampling, FFT and harmonic summing. We then scaled those results by the total number of beams in hilat ($N_{\mathrm{beams}} = 443287$) to evaluate the total cost of a complete search of the survey (Table \ref{tab:total_search_cost}). Although faster single-beam run times could be obtained for the old pipeline by allocating it more CPUs, this would translate into more total CPU-hours used in a full-scale search.

From these results it is clear that using the GPU-based \textsc{peasoup} pipeline made an acceleration search of HTRU possible with the specific resources we had at our disposal, running Fourier searching between 1.5 and 2 orders of magnitude faster than on CPUs. We note however that contrary to \textsc{peasoup}, \texttt{seek} was not designed to efficiently process multiple acceleration trials concurrently. This large speedup factor also created an interesting situation where folding candidates takes more time than a wide-range acceleration search.

\begin{table*}
\centering
\caption{Speed comparison between major steps of the old and new processing pipelines on a single hilat beam with a 270-second integration time. Benchmarks were performed on the Green II supercomputer, assuming an average of 650 DM trials, 30 acceleration trials and 50 folded candidates produced per beam. See \ref{subsec:speed_comparison} for details. It is interesting to note that folded candidate production could now become the main bottleneck of pulsar searching in some use cases at least.}
\label{tab:pipeline_runtimes}
\begin{tabular}{lllll}
\hline
\hline
                     & Operation      & Old Pipeline   & New Pipeline   &  Speedup factor \\
                     &                & (seconds)      & (seconds)      &   \\
\hline
Periodicity search   & Dedispersion   & 380            & 28             &  14 \\
                     & Search         & 5577           & 83             &  67 \\
                     & Sub-total      & 5957           & 111            &  54 \\        
\hline
Candidate production & Folding        & 142            & 53             &  2.7 \\
                     & RFI mitigation & 5              & 40             &  0.12 \\
                     & Optimization   & 97             & 65             &  1.5 \\  
                     & Sub-total      & 243            & 158            &  1.5 \\ 
\hline
TOTAL                &                & 6200           & 269            &  23.1 \\
\hline
\end{tabular}
\end{table*}

\begin{table}
\centering
\caption{Estimated resource consumption of a complete hilat acceleration search with both the old and new pipelines, based on the run times given in table \ref{tab:pipeline_runtimes}.}
\label{tab:total_search_cost}
\begin{tabular}{lll}
\hline
\hline
          & Old Pipeline & New Pipeline \\
\hline
CPU-hours & 2,300,000    & 58,000       \\
GPU-hours & 0            & 41,000       \\
\hline
\end{tabular}
\end{table}

\subsection{Output candidate quality and search completeness}

An in-depth output comparison between the old and new pipelines on the hilat survey section has been performed in Chapter 5 of \citet{CooperThesis}, of which we briefly summarize the results relevant to this section. This study was limited to slow pulsars with periods longer than 100 ms and our discussion will carry the same restrictions. Of interest is the survey area processed by both pipelines, called the overlap region hereafter and which contains 10,427 pointings, amounting to 31\% of the survey and nearly all the data that had been searched prior to this work. It was found that 42 known pulsars were detected by either pipeline in this overlap region; successful detection means that an associated candidate plot has been produced and is identifiable as a pulsar by eye. All 42 were detected by the new pipeline versus 38 for the old one. Furthermore, 3 of the new discoveries lying in the overlap region were not detected by the old pipeline either: PSR J0753$-$0816, PSR J1942$+$0147 and PSR J2228$-$65. The parameters of all 7 pulsars are summarized in Table \ref{tab:undetected_known_pulsars}. In at least two cases (PSR J0536$-$7543 and PSR J2048$-$1616), the old pipeline produced a candidate plot where the pulsar could not be identified due to the presence of RFI.

With small sample caveats in mind, the fact that 3 slow pulsar discoveries were not detected by the old pipeline over 10,427 pointings can be taken to be very significant: linearly scaled to the amount of data we searched (27,588 pointings), this would amount to an expected 8 pulsars missed. Considering that the current slow pulsar yield of hilat stands at 22 including the 11 new ones reported here (Table \ref{tab:discovery_params}), it means that changes in RFI mitigation schemes and other low-level processing steps have the potential to increase the slow pulsar yield by \textit{tens of percent}.

This validates the efficiency of the RFI mitigation algorithm presented in section \ref{sec:candidate_folding}, in real large-scale search conditions: no pulsars were lost due to its use and extra discoveries were enabled. But more importantly, this comparison underlines the large impact that various data manipulations (RFI mitigation being an important one, but not the only one) in the processing chain can have on search completeness before candidates even have a chance to reach the classification stage. We discuss this further in the next section.

\begin{table}
\centering
\caption{Parameters of the 7 pulsars \textit{not} detected by the old pipeline in hilat. Four were previously known, three are part of the new discoveries (Table \ref{tab:discovery_params}) and are marked with a star symbol. Their candidate plots were either not produced, or not recognizable. $D_{\mathrm{beam}}$ is the beam offset from the true source position in arcmin, and the S/N is that reported by the new pipeline.}
\label{tab:undetected_known_pulsars}
\begin{tabular}{lrrrrr}
\hline
\hline
PSR J         & {$P_0$}     & DM             & {$S_{1400}$} & {$D_{\mathrm{beam}}$} & S/N \\
              & {(s)}       & (pc cm$^{-3}$) &  {(mJy)}       & {(')}          &     \\
\hline
J0536$-$7543  & 1.246       & 18             &  13.0        &  12.3         & 33    \\
J0753$-$0816$^{*}$  & 2.093       & 38             &  $-$         &  $-$            & 15    \\
J1332$-$3032  & 0.650       & 15             &  0.30        &  7.1          & 17    \\
J1714$-$1054  & 0.696       & 51             &  $-$         &  7.0          & 18    \\
J1942$+$0147$^{*}$  & 1.405       & 151            &  $-$         &  $-$            & 15    \\
J2048$-$1616  & 1.962       & 12             &  13.0        &  9.0          & 366   \\
J2228$-$65$^{*}$    & 2.746       & 36             &  $-$         &  $-$            & 31    \\
\hline
\end{tabular}
\end{table}

\subsection{Candidate selection accuracy and human effort}

Two methods have been previously used to deal with candidate selection in HTRU med- and hilat searches: interactive candidate selection software such as \textsc{jreaper} \citep{Keith2009}, and another artificial neural network \citep[ANN; ][]{Bates2012}. With the first method, users project a batch of candidates on scatter plots where the axes are typically period and signal-to-noise ratio, or period and DM; from there they make a reduced selection of candidate plots to view in detail, exploiting tendencies of RFI candidates to cluster in narrow ranges of periods or around a dispersion measure of zero \citep{LevinThesis}. Due to its subjectiveness, the accuracy of this pre-selection process cannot be measured, precluding any comparison with ML algorithms. In terms of human effort, the total time spent viewing candidates to complete the initial medlat search using interactive software was approximately 100 person-hours.

The ANN of \citet{Bates2012} provides a more interesting point of discussion. It was both trained and evaluated on medlat candidates produced with the previous \textsc{sigproc} pipeline, and was found to have a false positive rate (FPR) of 0.3\% but a recall limited to 85\% (70\% on millisecond pulsars). Such an FPR is low enough to make the inspection of the entire survey output easily manageable by a single person, but a recall rate of 85\% offers a more limited discovery potential. We managed to significantly improve on the accuracy figures above, mainly due to the larger amount of training data we had at our disposal. \citet{Bates2012} trained their classifier soon after survey observations first started, on a limited data sample containing 70 candidates of the positive (pulsar) class. We were fortunate to have access to the whole survey and more than 1,000 such data instances, including discoveries from previous searches. They also used more candidate features (22 in total) in spite of having less training examples, which may have had a negative impact on accuracy (\S \ref{subsec:classification}).

%%%%%%%%%%%%%%%%%%%%%%%%%%%%%%%%%%%%%%%%%%%%%%%%%%%%%%%%%%%%%%%%%%%%%%%%%%%%%%%%%%%%%%%
% DISCUSSION SECTION
%%%%%%%%%%%%%%%%%%%%%%%%%%%%%%%%%%%%%%%%%%%%%%%%%%%%%%%%%%%%%%%%%%%%%%%%%%%%%%%%%%%%%%%
\section{Discussion and Conclusions}

In this paper we have described a pipeline with the two main functional elements that will be needed for future searches: a fast search code that leverages many-core architectures and an accurate candidate classification ML algorithm. We also devised and applied a new RFI mitigation method based on a simple outlier detection algorithm, which operates on folded data and has just one free parameter. The pipeline was used to search the intermediate and high latitude regions of the HTRU survey up to an acceleration of $\mathrm{50~m s^{-2}}$ and 23 new pulsars were discovered, including 6 MSPs and at least 4 in binary systems. The new search described here is shown to exceed the performance of previous searches on the same data both in terms of speed and completeness, and requires a much lower amount of human involvement in the candidate selection process. The improved completeness is evidenced by the fact that the majority of the new discoveries are isolated, slow pulsars that lie in a previously searched area of the parameter space.

The acceleration search of medlat and hilat was made possible by moving the time-domain re-sampling and Fourier search operations to the GPU, as they were by far the most time-consuming in the previous CPU-based pipeline. The resulting search code \textsc{peasoup} performs those tasks on order of 50 times faster (Table \ref{tab:pipeline_runtimes}), making candidate folding the new bottleneck. Speeding up candidate folding will likely become a requirement in the future, considering that, for example, the folding of up to 1000 candidates per beam is currently being considered for SKA pulsar searches \citep{SKAPulsarSearchesLevin2017}. GPUs are also a promising solution here. The folding of high time resolution spectra into a small number of phase bins is equivalent to histogram computation, and optimizing this operation on GPUs is an area of active research in computer science \citep[e.g.][]{Nugteren2011, GomezLuna2013}. Large speedups over CPU implementations have already been demonstrated.

The most important lesson learned from the hilat search is the large impact of RFI mitigation on search completeness, and the need to invest more effort into it. Since their introduction less than a decade ago \citep{Eatough2010}, candidate classification algorithms have greatly improved in accuracy, almost enough to be used on surveys envisioned with the 1500-beam SKA1-Mid \citep{SKAPulsarScienceKeane2017}. SPINN v2 would still have made the sifting of a 1500-beam version of HTRU med- and hilat doable within a few person-days, if a recall of 90\% can be considered acceptable \citep[Fig 5.1 of][]{MorelloMScT}. Lower error rates have since been achieved using ensemble classification \citep{Yao2016}. While this is welcome news, investing further effort in pushing classifiers to higher levels of accuracy is not the most cost-effective avenue if more than 10\% of potential discoveries can be lost in earlier stages of the processing chain.

We have pointed out the large potential impact of RFI mitigation on search completeness, but major concerns have also been raised about red noise removal schemes negatively affecting the detectability of pulsar signals \citep{Lazarus2015, VH2017}. Other processing steps such as harmonic summing, clustering of Fourier detections into candidates, or the time-domain candidate optimization process also have their own free parameters and the values chosen for these parameters impact search completeness in a way that is not quantitatively known. If one is ever to evaluate the recall and false positive rate of a pipeline \textit{as a whole} then it must be extensively tested on data generated by injecting simulated pulsar signals in randomly selected survey observations. Setting up such an evaluation framework may be computationally challenging but it would allow one to submit all internal pipeline parameters to an optimization process, similar to the training of a Machine Learning algorithm where its internal weights are fit to a sample of training data. And with a wide variety of generated pulsar signals available, candidate classifiers could be trained with nearly unlimited amounts of data, and the search completeness could be accurately known as a function of many physical pulsar parameters. Similar testing frameworks are already employed on the LIGO and Virgo gravitational wave detectors \citep[e.g.][]{LIGOSoftwareInjection, Biwer2017} to characterize detections caused by non-Gaussian noise, and one is planned to monitor the sensitivity of the CHIME FRB search \citep{CHIMEFRB}. Artificial signal injection has also been used as part of the PALFA survey to better evaluate the sensitivity of periodicity searches \citep{Lazarus2015, Parent2018}. As we move towards massively multibeam systems and real-time pulsar searches where the raw data are immediately discarded, with the additional requirement of automating the discovery process, pulsar searching pipelines are in danger of becoming completely opaque unless their whole processing chain is tested thoroughly.

\section*{Acknowledgements}

VM and BWS acknowledge funding from the European Research Council (ERC) under the European Union's Horizon 2020 research and innovation programme (grant agreement No. 694745). The Parkes Observatory is part of the Australia Telescope which is funded by the Commonwealth of Australia for operation as a National Facility managed by CSIRO. This work was supported by the Australian Research Council Centre for Excellence for All-sky Astrophysics (CAASTRO), through project number CE110001020. The re-processing of the HTRU survey made extensive use of the GPU Supercomputer for Theoretical Astrophysics Research (gSTAR) hosted at Swinburne University, and funded by a grant obtained via Astronomy Australia Limited (AAL). We extend our gratitude to Craig Heinke for greatly helping the writing of the Swift proposal for the confirmation attempt of J1618$-$36, and for performing the subsequent data analysis. Finally, we thank the referee Scott Ransom for providing valuable comments.



\bibliographystyle{mnras}
\bibliography{bib}

\begin{thebibliography}{}
\makeatletter
\relax
\def\mn@urlcharsother{\let\do\@makeother \do\$\do\&\do\#\do\^\do\_\do\%\do\~}
\def\mn@doi{\begingroup\mn@urlcharsother \@ifnextchar [ {\mn@doi@}
  {\mn@doi@[]}}
\def\mn@doi@[#1]#2{\def\@tempa{#1}\ifx\@tempa\@empty \href
  {http://dx.doi.org/#2} {doi:#2}\else \href {http://dx.doi.org/#2} {#1}\fi
  \endgroup}
\def\mn@eprint#1#2{\mn@eprint@#1:#2::\@nil}
\def\mn@eprint@arXiv#1{\href {http://arxiv.org/abs/#1} {{\tt arXiv:#1}}}
\def\mn@eprint@dblp#1{\href {http://dblp.uni-trier.de/rec/bibtex/#1.xml}
  {dblp:#1}}
\def\mn@eprint@#1:#2:#3:#4\@nil{\def\@tempa {#1}\def\@tempb {#2}\def\@tempc
  {#3}\ifx \@tempc \@empty \let \@tempc \@tempb \let \@tempb \@tempa \fi \ifx
  \@tempb \@empty \def\@tempb {arXiv}\fi \@ifundefined
  {mn@eprint@\@tempb}{\@tempb:\@tempc}{\expandafter \expandafter \csname
  mn@eprint@\@tempb\endcsname \expandafter{\@tempc}}}

\bibitem[\protect\citeauthoryear{{Abbott} et~al.,}{{Abbott}
  et~al.}{2016}]{LIGOSoftwareInjection}
{Abbott} B.~P.,  et~al., 2016, \mn@doi [Classical and Quantum Gravity]
  {10.1088/0264-9381/33/13/134001}, \href
  {http://adsabs.harvard.edu/abs/2016CQGra..33m4001A} {33, 134001}

\bibitem[\protect\citeauthoryear{{Barsdell}, {Bailes}, {Barnes}  \&
  {Fluke}}{{Barsdell} et~al.}{2012}]{Barsdell2012}
{Barsdell} B.~R.,  {Bailes} M.,  {Barnes} D.~G.,   {Fluke} C.~J.,  2012,
  \mn@doi [\mnras] {10.1111/j.1365-2966.2012.20622.x}, \href
  {http://adsabs.harvard.edu/abs/2012MNRAS.422..379B} {422, 379}

\bibitem[\protect\citeauthoryear{{Bates} et~al.,}{{Bates}
  et~al.}{2012}]{Bates2012}
{Bates} S.~D.,  et~al., 2012, \mn@doi [\mnras]
  {10.1111/j.1365-2966.2012.22042.x}, \href
  {http://adsabs.harvard.edu/abs/2012MNRAS.427.1052B} {427, 1052}

\bibitem[\protect\citeauthoryear{{Bhattacharyya} et~al.,}{{Bhattacharyya}
  et~al.}{2013}]{Bhattacharyya2013}
{Bhattacharyya} B.,  et~al., 2013, \mn@doi [\apjl]
  {10.1088/2041-8205/773/1/L12}, \href
  {http://adsabs.harvard.edu/abs/2013ApJ...773L..12B} {773, L12}

\bibitem[\protect\citeauthoryear{{Biwer} et~al.,}{{Biwer}
  et~al.}{2017}]{Biwer2017}
{Biwer} C.,  et~al., 2017, \mn@doi [\prd] {10.1103/PhysRevD.95.062002}, \href
  {http://adsabs.harvard.edu/abs/2017PhRvD..95f2002B} {95, 062002}

\bibitem[\protect\citeauthoryear{Breiman}{Breiman}{2001}]{Breiman2001}
Breiman L.,  2001, \mn@doi [Machine Learning] {10.1023/A:1010933404324}, 45, 5

\bibitem[\protect\citeauthoryear{{Burke-Spolaor} et~al.,}{{Burke-Spolaor}
  et~al.}{2011}]{BurkeSpolaor2011}
{Burke-Spolaor} S.,  et~al., 2011, \mn@doi [\mnras]
  {10.1111/j.1365-2966.2011.18521.x}, \href
  {http://adsabs.harvard.edu/abs/2011MNRAS.416.2465B} {416, 2465}

\bibitem[\protect\citeauthoryear{{Campana} \& {Di Salvo}}{{Campana} \& {Di
  Salvo}}{2018}]{Campana2018}
{Campana} S.,  {Di Salvo} T.,  2018, preprint, \href
  {http://adsabs.harvard.edu/abs/2018arXiv180403422C} {} (\mn@eprint {arXiv}
  {1804.03422})

\bibitem[\protect\citeauthoryear{Chandola, Banerjee  \& Kumar}{Chandola
  et~al.}{2009}]{Chandola2009}
Chandola V.,  Banerjee A.,   Kumar V.,  2009, \mn@doi [ACM Comput. Surv.]
  {10.1145/1541880.1541882}, 41, 15:1

\bibitem[\protect\citeauthoryear{{Cooper}}{{Cooper}}{2017}]{CooperThesis}
{Cooper} S.,  2017, PhD thesis, {The University of Manchester}

\bibitem[\protect\citeauthoryear{{Cordes} \& {Lazio}}{{Cordes} \&
  {Lazio}}{2002}]{NE2001}
{Cordes} J.~M.,  {Lazio} T.~J.~W.,  2002, ArXiv Astrophysics e-prints, \href
  {http://adsabs.harvard.edu/abs/2002astro.ph..7156C} {}

\bibitem[\protect\citeauthoryear{{Eatough}, {Molkenthin}, {Kramer}, {Noutsos},
  {Keith}, {Stappers}  \& {Lyne}}{{Eatough} et~al.}{2010}]{Eatough2010}
{Eatough} R.~P.,  {Molkenthin} N.,  {Kramer} M.,  {Noutsos} A.,  {Keith} M.~J.,
   {Stappers} B.~W.,   {Lyne} A.~G.,  2010, \mn@doi [\mnras]
  {10.1111/j.1365-2966.2010.17082.x}, \href
  {http://adsabs.harvard.edu/abs/2010MNRAS.407.2443E} {407, 2443}

\bibitem[\protect\citeauthoryear{{Eatough}, {Kramer}, {Lyne}  \&
  {Keith}}{{Eatough} et~al.}{2013}]{EatoughPMPS2013}
{Eatough} R.~P.,  {Kramer} M.,  {Lyne} A.~G.,   {Keith} M.~J.,  2013, \mn@doi
  [\mnras] {10.1093/mnras/stt161}, \href
  {http://adsabs.harvard.edu/abs/2013MNRAS.431..292E} {431, 292}

\bibitem[\protect\citeauthoryear{{Fruchter}, {Gunn}, {Lauer}  \&
  {Dressler}}{{Fruchter} et~al.}{1988}]{Fruchter1988b}
{Fruchter} A.~S.,  {Gunn} J.~E.,  {Lauer} T.~R.,   {Dressler} A.,  1988,
  \mn@doi [\nat] {10.1038/334686a0}, \href
  {http://adsabs.harvard.edu/abs/1988Natur.334..686F} {334, 686}

\bibitem[\protect\citeauthoryear{G\'omez-Luna, Gonz\'alez-Linares,
  Ignacio~Benavides  \& Guil}{G\'omez-Luna et~al.}{2013}]{GomezLuna2013}
G\'omez-Luna J.,  Gonz\'alez-Linares J.,  Ignacio~Benavides J.,   Guil N.,
  2013, Machine Vision and Applications, 24, 899

\bibitem[\protect\citeauthoryear{{Hastie}, {Tibshirani}  \&
  {Friedman}}{{Hastie} et~al.}{2009}]{StatLearning}
{Hastie} T.,  {Tibshirani} R.,   {Friedman} J.,  2009, The elements of
  statistical learning: data mining, inference and prediction, 2 edn.
Springer

\bibitem[\protect\citeauthoryear{{Hessels}, {Ransom}, {Stairs}, {Freire},
  {Kaspi}  \& {Camilo}}{{Hessels} et~al.}{2006}]{Hessels2006}
{Hessels} J.~W.~T.,  {Ransom} S.~M.,  {Stairs} I.~H.,  {Freire} P.~C.~C.,
  {Kaspi} V.~M.,   {Camilo} F.,  2006, \mn@doi [Science]
  {10.1126/science.1123430}, \href
  {http://adsabs.harvard.edu/abs/2006Sci...311.1901H} {311, 1901}

\bibitem[\protect\citeauthoryear{{Hessels} et~al.,}{{Hessels}
  et~al.}{2011}]{Hessels2011}
{Hessels} J.~W.~T.,  et~al., 2011, in {Burgay} M.,  {D'Amico} N.,  {Esposito}
  P.,  {Pellizzoni} A.,   {Possenti} A.,  eds,  American Institute of Physics
  Conference Series Vol. 1357, American Institute of Physics Conference Series.
  pp 40--43 (\mn@eprint {arXiv} {1101.1742}), \mn@doi{10.1063/1.3615072}

\bibitem[\protect\citeauthoryear{{Hotan}, {van Straten}  \&
  {Manchester}}{{Hotan} et~al.}{2004}]{Hotan2004}
{Hotan} A.~W.,  {van Straten} W.,   {Manchester} R.~N.,  2004, \mn@doi [\pasa]
  {10.1071/AS04022}, \href {http://adsabs.harvard.edu/abs/2004PASA...21..302H}
  {21, 302}

\bibitem[\protect\citeauthoryear{Johnston \& Kulkarni}{Johnston \&
  Kulkarni}{1991}]{jk91}
Johnston H.~M.,  Kulkarni S.~R.,  1991, ApJ, 368, 504

\bibitem[\protect\citeauthoryear{{Keane}}{{Keane}}{2017}]{SKAPulsarScienceKeane2017}
{Keane} E.~F.,  2017, preprint, \href
  {http://adsabs.harvard.edu/abs/2017arXiv171101910K} {} (\mn@eprint {arXiv}
  {1711.01910})

\bibitem[\protect\citeauthoryear{{Keane}, {Ludovici}, {Eatough}, {Kramer},
  {Lyne}, {McLaughlin}  \& {Stappers}}{{Keane} et~al.}{2010}]{KeanePMPS2010}
{Keane} E.~F.,  {Ludovici} D.~A.,  {Eatough} R.~P.,  {Kramer} M.,  {Lyne}
  A.~G.,  {McLaughlin} M.~A.,   {Stappers} B.~W.,  2010, \mn@doi [\mnras]
  {10.1111/j.1365-2966.2009.15693.x}, \href
  {http://adsabs.harvard.edu/abs/2010MNRAS.401.1057K} {401, 1057}

\bibitem[\protect\citeauthoryear{{Keane} et~al.,}{{Keane}
  et~al.}{2015}]{PulsarScienceSKA}
{Keane} E.,  et~al., 2015, Advancing Astrophysics with the Square Kilometre
  Array (AASKA14), \href {http://adsabs.harvard.edu/abs/2015aska.confE..40K}
  {p.~40}

\bibitem[\protect\citeauthoryear{{Keane} et~al.,}{{Keane}
  et~al.}{2018}]{SUPERBI}
{Keane} E.~F.,  et~al., 2018, \mn@doi [MNRAS] {10.1093/mnras/stx2126}, \href
  {http://adsabs.harvard.edu/abs/2018MNRAS.473..116K} {473, 116}

\bibitem[\protect\citeauthoryear{{Keith}, {Eatough}, {Lyne}, {Kramer},
  {Possenti}, {Camilo}  \& {Manchester}}{{Keith} et~al.}{2009}]{Keith2009}
{Keith} M.~J.,  {Eatough} R.~P.,  {Lyne} A.~G.,  {Kramer} M.,  {Possenti} A.,
  {Camilo} F.,   {Manchester} R.~N.,  2009, \mn@doi [\mnras]
  {10.1111/j.1365-2966.2009.14543.x}, \href
  {http://adsabs.harvard.edu/abs/2009MNRAS.395..837K} {395, 837}

\bibitem[\protect\citeauthoryear{{Keith} et~al.,}{{Keith} et~al.}{2010}]{HTRUI}
{Keith} M.~J.,  et~al., 2010, \mn@doi [\mnras]
  {10.1111/j.1365-2966.2010.17325.x}, \href
  {http://adsabs.harvard.edu/abs/2010MNRAS.409..619K} {409, 619}

\bibitem[\protect\citeauthoryear{{Knispel} et~al.,}{{Knispel}
  et~al.}{2013}]{Knispel2013}
{Knispel} B.,  et~al., 2013, \mn@doi [\apj] {10.1088/0004-637X/774/2/93}, \href
  {http://adsabs.harvard.edu/abs/2013ApJ...774...93K} {774, 93}

\bibitem[\protect\citeauthoryear{{Kocz}, {Bailes}, {Barnes}, {Burke-Spolaor}
  \& {Levin}}{{Kocz} et~al.}{2012}]{kbb+12}
{Kocz} J.,  {Bailes} M.,  {Barnes} D.,  {Burke-Spolaor} S.,   {Levin} L.,
  2012, \mn@doi [\mnras] {10.1111/j.1365-2966.2011.20029.x}, \href
  {http://esoads.eso.org/abs/2012MNRAS.420..271K} {420, 271}

\bibitem[\protect\citeauthoryear{{Lazarus}}{{Lazarus}}{2013}]{Lazarus2013}
{Lazarus} P.,  2013, in {van Leeuwen} J.,  ed.,  IAU Symposium Vol. 291,
  Neutron Stars and Pulsars: Challenges and Opportunities after 80 years. pp
  35--40 (\mn@eprint {arXiv} {1210.8003}), \mn@doi{10.1017/S1743921312023101}

\bibitem[\protect\citeauthoryear{{Lazarus} et~al.,}{{Lazarus}
  et~al.}{2015}]{Lazarus2015}
{Lazarus} P.,  et~al., 2015, \mn@doi [\apj] {10.1088/0004-637X/812/1/81}, \href
  {http://adsabs.harvard.edu/abs/2015ApJ...812...81L} {812, 81}

\bibitem[\protect\citeauthoryear{{Levin}}{{Levin}}{2012}]{LevinThesis}
{Levin} L.,  2012, PhD thesis, {Swinburne University of Technology}, \url
  {http://hdl.handle.net/1959.3/312190}

\bibitem[\protect\citeauthoryear{{Levin} et~al.,}{{Levin}
  et~al.}{2017}]{SKAPulsarSearchesLevin2017}
{Levin} L.,  et~al., 2017, preprint, \href
  {http://adsabs.harvard.edu/abs/2017arXiv171201008L} {} (\mn@eprint {arXiv}
  {1712.01008})

\bibitem[\protect\citeauthoryear{{Lorimer}, {Bailes}, {McLaughlin}, {Narkevic}
  \& {Crawford}}{{Lorimer} et~al.}{2007}]{LorimerBurst}
{Lorimer} D.~R.,  {Bailes} M.,  {McLaughlin} M.~A.,  {Narkevic} D.~J.,
  {Crawford} F.,  2007, \mn@doi [Science] {10.1126/science.1147532}, \href
  {http://adsabs.harvard.edu/abs/2007Sci...318..777L} {318, 777}

\bibitem[\protect\citeauthoryear{{Lyne}, {Stappers}, {Keith}, {Ray}, {Kerr},
  {Camilo}  \& {Johnson}}{{Lyne} et~al.}{2015}]{Lyne2015}
{Lyne} A.~G.,  {Stappers} B.~W.,  {Keith} M.~J.,  {Ray} P.~S.,  {Kerr} M.,
  {Camilo} F.,   {Johnson} T.~J.,  2015, \mn@doi [\mnras]
  {10.1093/mnras/stv236}, \href
  {http://adsabs.harvard.edu/abs/2015MNRAS.451..581L} {451, 581}

\bibitem[\protect\citeauthoryear{{Lyon}, {Stappers}, {Cooper}, {Brooke}  \&
  {Knowles}}{{Lyon} et~al.}{2016}]{Lyon2016}
{Lyon} R.~J.,  {Stappers} B.~W.,  {Cooper} S.,  {Brooke} J.~M.,   {Knowles}
  J.~D.,  2016, \mn@doi [\mnras] {10.1093/mnras/stw656}, \href
  {http://adsabs.harvard.edu/abs/2016MNRAS.459.1104L} {459}

\bibitem[\protect\citeauthoryear{{Manchester} et~al.,}{{Manchester}
  et~al.}{2001}]{Manchester2001}
{Manchester} R.~N.,  et~al., 2001, \mn@doi [\mnras]
  {10.1046/j.1365-8711.2001.04751.x}, \href
  {http://adsabs.harvard.edu/abs/2001MNRAS.328...17M} {328, 17}

\bibitem[\protect\citeauthoryear{{Manchester}, {Hobbs}, {Teoh}  \&
  {Hobbs}}{{Manchester} et~al.}{2005}]{PSRCAT}
{Manchester} R.~N.,  {Hobbs} G.~B.,  {Teoh} A.,   {Hobbs} M.,  2005, \mn@doi
  [\aj] {10.1086/428488}, \href
  {http://cdsads.u-strasbg.fr/abs/2005AJ....129.1993M} {129, 1993}

\bibitem[\protect\citeauthoryear{{Morello}}{{Morello}}{2016}]{MorelloMScT}
{Morello} V.,  2016, Master's thesis, {Swinburne University of Technology},
  \url {http://hdl.handle.net/1959.3/434704}

\bibitem[\protect\citeauthoryear{{Morello}, {Barr}, {Bailes}, {Flynn}, {Keane}
  \& {van Straten}}{{Morello} et~al.}{2014}]{Morello2014}
{Morello} V.,  {Barr} E.~D.,  {Bailes} M.,  {Flynn} C.~M.,  {Keane} E.~F.,
  {van Straten} W.,  2014, \mn@doi [\mnras] {10.1093/mnras/stu1188}, \href
  {http://adsabs.harvard.edu/abs/2014MNRAS.443.1651M} {443, 1651}

\bibitem[\protect\citeauthoryear{{Ng}}{{Ng}}{2017}]{PulsarScienceCHIME}
{Ng} C.,  2017, preprint, \href
  {http://adsabs.harvard.edu/abs/2017arXiv171102104N} {} (\mn@eprint {arXiv}
  {1711.02104})

\bibitem[\protect\citeauthoryear{{Ng} et~al.,}{{Ng}
  et~al.}{2015}]{NgLowlat2015}
{Ng} C.,  et~al., 2015, \mn@doi [\mnras] {10.1093/mnras/stv753}, \href
  {http://adsabs.harvard.edu/abs/2015MNRAS.450.2922N} {450, 2922}

\bibitem[\protect\citeauthoryear{Nugteren, van~den Braak, Corporaal  \&
  Mesman}{Nugteren et~al.}{2011}]{Nugteren2011}
Nugteren C.,  van~den Braak G.-J.,  Corporaal H.,   Mesman B.,  2011, in
  Proceedings of the Fourth Workshop on General Purpose Processing on Graphics
  Processing Units. GPGPU-4.
ACM, New York, NY, USA, pp 1:1--1:8, \mn@doi{10.1145/1964179.1964181}, \url
  {http://doi.acm.org/10.1145/1964179.1964181}

\bibitem[\protect\citeauthoryear{{Parent} et~al.,}{{Parent}
  et~al.}{2018}]{Parent2018}
{Parent} E.,  et~al., 2018, \mn@doi [\apj] {10.3847/1538-4357/aac5f0}, \href
  {http://cdsads.u-strasbg.fr/abs/2018ApJ...861...44P} {861, 44}

\bibitem[\protect\citeauthoryear{{Polzin} et~al.,}{{Polzin}
  et~al.}{2018}]{Polzin2018}
{Polzin} E.~J.,  et~al., 2018, \mn@doi [\mnras] {10.1093/mnras/sty349}, \href
  {http://adsabs.harvard.edu/abs/2018MNRAS.476.1968P} {476, 1968}

\bibitem[\protect\citeauthoryear{{Ransom}, {Eikenberry}  \&
  {Middleditch}}{{Ransom} et~al.}{2002}]{Ransom2002}
{Ransom} S.~M.,  {Eikenberry} S.~S.,   {Middleditch} J.,  2002, \mn@doi [\aj]
  {10.1086/342285}, \href {http://adsabs.harvard.edu/abs/2002AJ....124.1788R}
  {124, 1788}

\bibitem[\protect\citeauthoryear{{Stappers}, {Bailes}, {Lyne}, {Camilo},
  {Manchester}, {Sandhu}, {Toscano}  \& {Bell}}{{Stappers}
  et~al.}{2001}]{Stappers2001}
{Stappers} B.~W.,  {Bailes} M.,  {Lyne} A.~G.,  {Camilo} F.,  {Manchester}
  R.~N.,  {Sandhu} J.~S.,  {Toscano} M.,   {Bell} J.~F.,  2001, \mn@doi
  [\mnras] {10.1046/j.1365-8711.2001.04074.x}, \href
  {http://adsabs.harvard.edu/abs/2001MNRAS.321..576S} {321, 576}

\bibitem[\protect\citeauthoryear{{Staveley-Smith} et~al.,}{{Staveley-Smith}
  et~al.}{1996}]{StaveleySmith1996}
{Staveley-Smith} L.,  et~al., 1996, \pasa, \href
  {http://adsabs.harvard.edu/abs/1996PASA...13..243S} {13, 243}

\bibitem[\protect\citeauthoryear{{The CHIME/FRB Collaboration} et~al.,}{{The
  CHIME/FRB Collaboration} et~al.}{2018}]{CHIMEFRB}
{The CHIME/FRB Collaboration} et~al., 2018, preprint, \href
  {http://adsabs.harvard.edu/abs/2018arXiv180311235T} {} (\mn@eprint {arXiv}
  {1803.11235})

\bibitem[\protect\citeauthoryear{{Thornton}}{{Thornton}}{2013}]{ThorntonThesis}
{Thornton} D.,  2013, PhD thesis, {The University of Manchester}, \url
  {https://www.escholar.manchester.ac.uk/uk-ac-man-scw:212510}

\bibitem[\protect\citeauthoryear{{Thornton} et~al.,}{{Thornton}
  et~al.}{2013}]{Thornton2013}
{Thornton} D.,  et~al., 2013, \mn@doi [Science] {10.1126/science.1236789},
  \href {http://adsabs.harvard.edu/abs/2013Sci...341...53T} {341, 53}

\bibitem[\protect\citeauthoryear{Tukey}{Tukey}{1977}]{Tukey1977}
Tukey J.~W.,  1977, Exploratory Data Analysis.
Addison-Wesley

\bibitem[\protect\citeauthoryear{Yao, Xin  \& Guo}{Yao et~al.}{2016}]{Yao2016}
Yao Y.,  Xin X.,   Guo P.,  2016, in 2016 12th International Conference on
  Computational Intelligence and Security (CIS). pp 120--124,
  \mn@doi{10.1109/CIS.2016.0036}

\bibitem[\protect\citeauthoryear{{van Heerden}, {Karastergiou}  \&
  {Roberts}}{{van Heerden} et~al.}{2017}]{VH2017}
{van Heerden} E.,  {Karastergiou} A.,   {Roberts} S.~J.,  2017, \mn@doi
  [\mnras] {10.1093/mnras/stw3068}, \href
  {http://adsabs.harvard.edu/abs/2017MNRAS.467.1661V} {467, 1661}

\bibitem[\protect\citeauthoryear{{van Straten} \& {Bailes}}{{van Straten} \&
  {Bailes}}{2011}]{DSPSR}
{van Straten} W.,  {Bailes} M.,  2011, \mn@doi [\pasa] {10.1071/AS10021}, \href
  {http://adsabs.harvard.edu/abs/2011PASA...28....1V} {28, 1}

\makeatother
\end{thebibliography}

%%%%%%%%%%%%%%%%%%%%%%%%%%%%%%%%%%%%%%%%%%%%%%%%%%

% Don't change these lines
\bsp	% typesetting comment
\label{lastpage}
\end{document}